\documentclass[]{aa}
\newcommand{\apropto}{\;
  \raise0.3ex\hbox{$\propto$\kern-0.75em\raise-1.1ex\hbox{$\sim$
  }}\;\hskip-2pt }
\usepackage{txfonts,graphicx,amssymb,subfigure}
\usepackage[normalem]{ulem}
\usepackage{xcolor}

\begin{document}
\title{Dynamo generated magnetic configurations in accretion discs and the nature of quasi-periodic oscillations in accreting binary systems}

   \author{D.~Moss\inst{1}
          \and
        D.~Sokoloff \inst{2}
          \and
         V.~Suleimanov\inst{3,4}
          }

   \offprints{D.~Moss, david.moss@manchester.ac.uk}

   \institute{ School of Mathematics, University of Manchester, Oxford Road, Manchester, M13 9PL, UK
   \and
    Department of Physics, Moscow University, 119992 Moscow, Russia
               \and
Institut f\"ur Astronomie und Astrophysik, Kepler Center for Astro and
Particle Physics,
Universit\"at T\"ubingen, Sand 1,
 72076 T\"ubingen, Germany
\and
Kazan (Volga region) Federal University, Kremlevskaja str., 18, Kazan 420008, Russia
    }

   \date{Received ..... ; accepted .....}

\abstract{Magnetic fields are important for accretion disc structure. Magnetic fields in a disc 
system may be transported with the accreted  matter.  
They can be associated with either  the central body and/or jet,
and be fossil or dynamo excited in situ.}
{We consider dynamo excitation of magnetic fields in accretion discs of accreting binary systems in an attempt to clarify 
possible configurations of dynamo generated magnetic fields. We first model the entire disc with realistic radial extent and thickness using an alpha-quenching non-linearity. We then study the simultaneous effect of feedback from the
Lorentz force from the dynamo-generated field.}
{We perform numerical simulations in the framework of a relatively simple mean-field model which allows the generation of global
magnetic configurations.}
{We explore a range
of possibilities for the dynamo number, and find quadrupolar-type 
solutions with irregular temporal oscillations that might be compared to observed rapid luminosity fluctuations. 
The dipolar symmetry models with $R_\alpha<0$  have lobes of strong toroidal field adjacent 
to the rotation axis that could be relevant to jet launching phenomena.}
{We have explored and extended the solutions known for thin accretion discs.}


\titlerunning{Magnetic fields in accretion discs}
\authorrunning{Moss et al.}

\maketitle

\section{Introduction}

Starting from the seminal paper of Shakura and Sunyaev (1973), magnetic fields have been seen  to play an  important 
role in various 
explanations of the accretion discs physics and phenomenology. In particular, 
magnetic fields are central to  
understanding of the angular momentum transport in the disc as well as the collimation by magnetic fields of jets in 
various accretion disc systems; see e.g. \cite{BH:91,BZ:77,BP:82,S.etal:96,NMcCT:14}.

More generally, accretion discs are discussed in several branches of astronomy and the associated magnetic 
fields can, in principle, have various natures. Broadly speaking, magnetic fields in a disc system may be transported 
with the accreted  matter, 
either being associated with the central body and/or jet,
and can be fossil or dynamo excited in situ  (e.g. Lubow et al. 1994, Okuzumi et al. 2014).

The important point is that a rotating accretion disc is a body which 
can also provide dynamo excitation. Indeed, accretion discs are thought to 
possess differential rotation and 
turbulence, i.e. two types of motions important for dynamo action. 
The medium of accretion discs is believed to be sufficiently 
conductive that induction effects dominate over dissipation. Coriolis force and density stratification
affect turbulence and are assumed to  make it mirror asymmetric. As a result, all the components of conventional 
disc dynamos that have been
thoroughly investigated for the discs of spiral galaxies are also found 
in accretion discs,  and it can be expected that
magnetic field excitation of a nature quite similar to that in spiral galaxies will occur.

On the other hand, the hydrodynamics of accretion discs is obviously 
different in some respects from that of 
spiral galaxies. In particular, flat rotation curves are typical of spiral galaxies whilst a Keplerian rotation law is
usually assumed for accretion discs.  
Moreover, 
substantial radial flows are thought to occur in accretion discs,
whereas radial flows appear to be weak in the greater part of most spiral galaxies.

Of course, the problem has attracted the attention of dynamo theorists,
 and a number of papers have been focused on
simulating dynamo processes in the framework of direct numerical simulations, 
typically in a sheared box representing a part 
of an accretion disc (e.g. Pariev \& Colgate, 2007a,b; Stone et al. 1996; Davis et al. 2010; Jiang et al. 2013). 
The message from these investigations is that 
dynamo action quite similar to that in
galactic discs also occurs  in accretion discs. In particular, Gressel \& Pessah (2015), based on the analysis of a quite 
sophisticated numerical model, conclude that their findings give 
additional support to the importance of the $\alpha 
\Omega$-mechanism in determining the evolution of large-scale magnetic fields 
in turbulent accretion disks; see also Blackman (2012). 

On the other hand, it is quite problematic to extrapolate a 
shear box simulation to a general scenario for magnetic field configurations 
in entire accretion discs, and to identify 
the link between the disc parameters and such configurations.  
Of course, it is very important to 
connect the local shear-box simulations with global disc dynamo action (e.g. Sadowski et al. 2014).
However, 
understanding the link between the physical parameters of the 
disc and the dynamo driven magnetic configuration is still problem that
requires clarification.
Certainly, it is not clear in advance how specific the configuration of dynamo generated magnetic 
field in accretion discs is. 

A common practice in studies of galactic discs is to consider relatively
simple mean-field models as a complement to direct numerical simulations.
Mean-field modelling provides an understanding of the magnetic 
configuration as a whole ignoring finer details. In our 
opinion, study of mean-field dynamo models for accretion discs can provide a valuable synergy with
direct numerical simulations in shear boxes.  This is the aim of the present paper. 
We begin by modelling the entire radial extent of the disc, with a realistic vertical 
thickness, for a range of dynamo parameters, using a quasi-kinematic approach 
(i.e. with a simple alpha-quenching non-linearity only). Then, with a slightly radially
truncated model we also include a representation of the action of the
Lorentz force of the large-scale field on the azimuthal velocities.

Our strategy is to consider as a basic example a standard hydrodynamical model for accretion discs of cataclysmic 
variables as celestial bodies which includes as few non-standard physics as possible, and to play with numbers 
around those of the basic model. 
We anticipate that this will provide hints as to what might be expected from 
such an approach applied to discs around black holes.
We concentrate on the disc dynamo only and for the time being ignore what 
might happen arising from the interplay between the disc and the central body. 
This approach was exploited in the 1990s 
as being the only one then affordable
(e.g. Stepinsky \& Levi 1990, Torkelsson \& Brandenburg  1994, Reyes-Ruiz \& Stepinski 1995,
R\"udiger, Elstner \& Stepinski 1995, among a number of others). The models studied in
these papers were constrained by the facilities then readily available to have either
unrealistically thick discs, constant thickness discs, or discs truncated 
at relatively  large inner radii. (Arlt \& R\"udiger 2001, among others,
 studied the role of magnetorotational instability -- MRI -- we assume that its effects
are broadly speaking subsumed into our turbulence coefficients.) 
At that time efforts were concentrated to some extent on 
the computational aspects of the problem and then on
the transition from mean-field models to direct numerical 
simulations (DNS) that soon became feasible (e.g. Brandenburg et al.  1995). 
On one hand, this seminal progress in direct 
numerical simulations clarified the physical nature of accretion disc dynamos,
but did not allow 
extraction of all 
the astronomically important results that were available
from simple mean-field models as required by the wider astronomical 
community. In this sense our paper is a revisiting of the older approach,
but can now be viewed in the light of the contemporary understanding of 
the relation between abilities of modern DNS 
to study the detailed physics and the helpful support of the more 
traditional tools of mean-field. 

We emphasize specifically that we will only consider magnetic fields generated by 
the dynamo in the accretion 
disc itself, rather in its
surroundings.

Our paper is mainly addressed to workers in the astronomical interpretation 
of accretion disc phenomena. Of 
course, experts in dynamo modelling know without specific simulations that the simplest dynamo model for a thin 
disc will have a steady magnetic configuration of quadrupole symmetry,
concentrated near the central plane of 
the disc. The problem however is that this idea has not fully 
penetrated the observational community,
and that the constraints arising need to be discussed. In particular, a naive 
generalization of experience from galactic dynamos to accretion discs is risky, and may be sometimes misleading because the nature of interstellar turbulence 
and that of turbulence in accretion discs is different. On
the other hand it seems preferable to keep in mind the rich mass of 
experince obtained from studies of dynamos in galactic discs,
 supported by direct observations, rather than just to ignore it.

\section{Disc model} 

The conventional scenario of magnetic field generation, the $\alpha \Omega$-dynamo, is based on the joint action of 
differential rotation and mirror-asymmetric turbulence. It is conventional to represent the relative importance of 
these effects in
terms of several dimensionless numbers similar to the Reynolds number in hydrodynamics. Our first aim is to present 
conservative estimates for these numbers in the discs of cataclysmic variables, and then discuss more-or-less 
realistic scattering of numbers around this estimates.

\subsection{Basic model}
\label{basic}

We consider the steady state optically thick and geometrically thin accretion $\alpha$-disc (Shakura \& Sunyaev 
1973) of a white dwarf.
The mass accretion rate $\dot M = 10^{17}$~g s$^{-1}$ was chosen as
being a typical value for nova-like variables and 
dwarf novae during outbursts
(Warner 2003). Both of these types of objects are sub-classes of cataclysmic variable stars.  We note, however, 
that 
our results will not change qualitatively 
if we take larger values of $\dot M$.
We take a typical white dwarf mass $M= 0.8 M_\odot = 1.6 \times 10^{33}$ g with radius $R_{0} = 7 \times  10^8$ cm, 
which
follows from the universal mass-radius relation for the white dwarfs (Nauenberg, 1972). The  outer disc radius is 
taken to be
$R_{\rm out}= 7 \times 10^{10}$ cm ($R_{\rm out}/R_0 \approx 10^2$). 
We assume that the central white dwarf is not (or weakly) magnetized, and so the
inner radius of the disc  
is close to that of the radius of the central white dwarf $R_{\rm in}=R_0$.
We use $R_{\rm out}$ as a length  unit, $r=R/R_{\rm out}$ so 
that in dimensionless units $r_{\rm out}=1$, $r_{\rm in} = 10^{-2}$.

The disc model is essentially that of Shakura \& Sunyaev (1973). 
One of the important problems for the accretion disc theory is a boundary condition at the inner disc radius. 
Standard $\alpha$-disc theory (Shakura \& Sunyaev 1973) for accretion discs 
around black holes uses as boundary condition the vanishing of the 
viscous stress tensor. This condition leads to
formally infinite or zero values of the basic physical properties at the inner 
disc radius.  This appears mathematically as a numerical factor 
$Q(r)=\left(1-\sqrt{\frac{r_{\rm in}}{r}}\right)$ applied
to various powers in the analytical 
solutions for the all the physical parameters. To avoid   these infinities we 
used below a modified factor
   $Q(r)=\left(1-q\sqrt{\frac{0.01}{r}}\right)$ with $q=0.9$.

In fact, the inner boundary condition problem is much more complicated for the accretion discs around white dwarfs. 
A boundary layer between a slowly rotating 
white dwarf and a fast rotating (with Keplerian angular velocity) accretion disc must exist (Pringle \& Savonije 
1979, Tylenda 1977, Popham \& Narayan 1995,
Hertfelder et al. 2013). The dynamo action in the boundary layers is a separate problem, which we neglect here,
but plan to consider in future work. Here we are interested in
magnetic field generation in the accretion disc itself, and the assumed approximation is enough for our aims.

The disc in the standard theory rotates  with  Keplerian velocity, so that

\begin{equation}
V = \sqrt{GM/R}, \quad \quad \Omega = \sqrt{GM/R^3}.
\label{rot}
\end{equation} 
Thus $V = 389\,\, r^{-1/2} $ km s$^{-1}$, $\Omega = 5.56\times 10^{-4}\,\,r^{-3/2}$~s$^{-1}$, and the rotation period is  
$P=2\pi \Omega^{-1}= 11\,298\,\,  r^{3/2}$ s.
It is worth noting that Eq.~(\ref{rot}) is quite different to the typical
rotation curves of spiral galaxies: the latter have almost flat $V(r)$, presumably because of the presence of dark 
matter halos around 
galactic discs.

To calculate all physical properties in the disc model described below   
we used the analytical solutions obtained
by Suleimanov et al. (2007) on the basis of the Shakura \& Sunyaev (1973) model, but with the  appropriate opacity 
for the solar mix plasma and the correct
vertical disc structure.  The solutions derived for the gas pressure and the absorption opacity dominated accretion 
discs are
valid for our disc model.  We used a theoretical turbulent viscosity 
parameter $\alpha_{\rm SS} = 0.1$. The radial distributions of our
mid-plane temperatures and density are
\begin{equation}
    T = 1.51\times 10^4\, r^{-3/4}\,Q(r)^{3/10}\,\,{\rm K},
\end{equation}
and
\begin{equation}
    \rho = 7.03\times 10^{-9}\, r^{-15/8}\,Q(r)^{11/20}\,\,{\rm g\, cm^{-3}}.
\end{equation}

A basic parameter for the mean-field magnetic dynamo model is a turbulent diffusivity $\eta = lV_{\rm t}/3$, where 
$l$ is a typical turbulence 
correlation length,
and $V_{\rm t}$ is an r.m.s. value of the turbulent velocities. Formally, we have to consider $\eta$ values that 
are consistent with the assumed $\alpha_{\rm SS}$,
but taking into account that the above estimate is certainly over simplistic we consider it as a free parameter in order to 
investigate various regimes of magnetic dynamo action the 
accretion disc.
 
Accretion discs are assumed to be turbulent, and a naive estimate for the
r.m.s. value of the turbulent velocities $V_{\rm t}$ is that it is a fraction of the sound speed, $V_{\rm s}$, $V_{\rm t} = \sigma_{\rm v} V_{\rm s}$, 
$\sigma_{\rm v} \le1$.
For the basic model we assume $\sigma_{\rm v} = 1$.
Therefore, the turbulent diffusivity is estimated by 
\begin{equation} \label{et}
\eta= \frac{l V_{\rm t}}{3} = \sigma_{\rm l}\,\sigma_{\rm v}\,\frac{H V_{\rm s}}{3}.
\end{equation}

In our approach the sound speed can be evaluated as
\begin{equation}
V_{\rm s} =  \left(\frac{k_{\rm B}T(r)}{\mu m_{\rm H}}\right)^{1/2} =  14.2\, r^{-3/8} Q(r)^{3/20} \, {\rm km\, s}^{-1}.
\end{equation}
Thus $V_{\rm s}$ has a maximum of 53 km s$^{-1}$ at $r=1.36 \, r_{\rm in}$ and 
smoothly decreases  to 
$V_{\rm s}= 14$ km s$^{-1}$ at $r=1$.

Usually it is assumed that $l$ is close to, although somewhat smaller than, the disc scale height $H$. To be 
specific we use $l= \sigma_{\rm l} H$. 
For the basic model we assume $\sigma_{\rm l} = 0.2$. Fortunately, 
results do not depend crucially on this estimate.We take the disc scale height $H$ as an initial estimate for the 
disc thickness
\begin{equation}
H= V_{\rm s}/\Omega = 2.55 \times 10^9 \,r^{9/8} Q(r)^{3/20} \, {\rm cm}.
\label{gasp}
\end{equation}
Thus the local aspect ratio is
$H/R \approx 10^{-2}$
\begin{equation}
\lambda(r) =\frac{H}{R} = 0.037\,  r^{1/8} \,Q(r)^{3/20}.
\end{equation}

The infall radial velocity of matter in the disc is
\begin{equation}
V_{\rm in}=\dot{M}/(2\pi R \,H\rho)= 0.13\, 
r^{-1/4}\, Q(r)^{-7/10}\,\,{\rm km\,s^{-1}}
\end{equation}

Another important physical parameter is the equipartition magnetic field 
strength, which can be estimated by assuming that
 the magnetic pressure is in equilibrium with the energy of the turbulent
gas motions 
\begin{equation}
    B_{\rm eq} = \sigma_{\rm v} V_{\rm s} \sqrt{4\pi\,\rho} = 420\,\sigma_{\rm v}\, r^{-21/16}\,Q(r)^{17/40}\,\,{\rm G}.
\label{Beq}
\end{equation}

Excitation of mean field disc dynamos is usually assumed to be by the 
$\alpha\,\Omega$
mechanism, based on the joint action of differential rotation $\Omega(r)$
and the mirror asymmetric (from the action of the Coriolis force) turbulence:
$\alpha$ is a measure of this asymmetry and has dimensions of velocity.
Conventionally 
\begin{equation} \label{al}
\alpha=l^2\,\frac{\Omega}{H} = \sigma_{\rm l}^2\, V_{\rm s}
\end{equation}
 in $z>0$ (and is antisymmetric w.r.t. $z$). 

Now we can define dimensionless measures of the contributions to dynamo
action of each from the principal drivers. Thus
\begin{equation} \label{rw}
R_\omega=\frac{H^2\Omega}{\eta}=V_{\rm s} \frac{H}{\eta} = \frac{3}{\sigma_{\rm l}\sigma_{\rm v}},
\end{equation}
 and 
\begin{equation} \label{ra}
 R_\alpha = \alpha\,\frac{H}{\eta} = R_\omega\,\frac{\alpha}{V_{\rm s}} =3 \frac{\sigma_{\rm l}}{\sigma_{\rm v}},
\end{equation}
where Eqs.\,(\ref{et}) and (\ref{al}) were used.
These determinations of $R_{\omega}$ and $R_\alpha$ give a local dynamo number
\begin{equation}
D(r)=R_\alpha R_\omega=\frac{H^3 \alpha\,\Omega}{\eta^2} = \alpha\,V_{\rm s} \frac{H^2}{\eta^2} = 9\, \sigma_{\rm v}^{-2}.
\label{dnum}
\end{equation}
Remarkably, $D(r)$ is independent of radius if we assume that $\sigma_{\rm v}$ is a constant over the disc.
(The numerical factor 3 in Eq.\,(\ref{et}) comes from the dimension of the 
space.) There is also a subsidiary parameter, a Reynolds number
\begin{equation}
Rm=V_{\rm in,0} \frac{H_0}{\eta_0} = R_\omega \,\frac{V_{\rm in,0}}{V_{\rm s,0}} \approx 0.14 \left(\frac{0.2}
{\sigma_{\rm v}\sigma_{\rm l}}\right),
\end{equation}
 which is found to play a minor role.

The above estimate of dynamo number $D$ is well-known in 
dynamo theory, and discussed by Stepinski 
\& Levi (1990). Our aim here is to emphasize the constraints that follow from 
this estimate in simple dynamo models,
as well as ways of proceeding further. In particular, already 
R\"udiger, Elstner \& Stepinski (1995)
mentioned that the estimate only determines properties of 1D disc models,
and that 2D models need separate, more detailed, consideration.

 It is also useful to define a magnetic diffusive time scale for the disc model
 \begin{equation} \label{ts}
 \tau = \frac{H^2}{\eta} = R_\omega\,\frac{H}{V_{\rm s}} =\frac{3}{\sigma_{\rm l}\,\sigma_{\rm v}} \Omega^{-1} 
 \approx 27\,780\,\left(\frac{0.2}{\sigma_{\rm l}\,\sigma_{\rm v}}\right)\,r^{3/2}\,{\rm s}.
\end{equation}

For comparison, the
dynamo number in the discs of spiral galaxies is also estimated to be about 
$D = 9$, and there results in excitation of 
large-scale magnetic fields of quadrupole symmetry that are almost totally 
confined to the disc. According to 
experience from discs of spiral galaxies, generation of 
magnetic fields of dipole symmetry in the space surrounding the disc requires $D > 100$ at 
least. Just for orientation, $D = 10^3 \sim 10^5$ in the spherical shell inside the Sun where dynamo action
resulting in generation of the solar dipole magnetic field is thought to occur.

Of course, the above estimates provides a hint only and specific modelling of
dynamo action in accretion discs
is essential. However we can conclude that the scaling of Eq.~(\ref{dnum}) 
plays a crucial role in the 
problem.
The underlying physics for the Eq.~(\ref{dnum}) is the identification of the turbulent velocities of the accretion 
disc medium and the sound speed, which in turn determines the pressure height scale of the disc $H$.

\subsection{Parameter space near the basic model}
\label{param1}

Given that our basic model certainly does not reproduce accurately
and in detail the properties of real accretion discs, we need to
consider plausible modifications of the model and thus of the estimate
(\ref{dnum}).
Our 'standard' case (above) has $\sigma_{\rm v}=1, \sigma_{\rm l}=0.2$.
Inter alia, we can adjust the relative roles of differential rotation
and mirror asymmetric turbulence rescaling $\sigma_{\rm v}$ and $\sigma_{\rm l}$ while keeping $D$ constant.

Using the above scaling we can obtain the dynamo number  $D \gg 9$, i.e. if the turbulent 
velocities are substantially lower then the sound speed ($\sigma_v \ll 1$). Therefore, the dynamo number can be 
substantially larger 
than that in the basic model and the variety of dynamo driven magnetic 
configuration may become richer. 
More specifically,  we need $\sigma_v = 0.1 \sim 0.3$ to expect something more than the quadrupolar field excitation 
that is well-known from galactic dynamo 
studies. Below (Sect.~4.4) we face one more problem with definition of dynamo governing parameters: when developing 
our model from 1D to 2D we have to modify the definitions of $R_\alpha$, $R_\omega$ and $R_m$. This can be done by a 
multiplier which formally plays the same role as $\sigma_v$ however its physical meaning is different. In order to 
emphasize this point we introduce a quantity $f_{\rm v}$ multiplying each of
 $R_\alpha, R_\omega$ and $R_m$ simultaneously. 
This allows a systematic exploration of parameter space. From a dynamo theory viewpoint, 
it is natural to test the sensitivity of the model to changes in dynamo number including the cases $f_{\rm v} <1$ as well 
as $f_{\rm v} >1$ while $\sigma_v >1$ ($f_{\rm v}<1$) formally corresponds to supersonic 
turbulence which taken literally would  not be realistic. In the restricted 
context of dynamo
theory this restriction does not apply, and in any case it would be rash
to assume that our model for the disc can be interpreted too exactly. 

Dynamo action for a given flow gives an exponentially growing magnetic field which quite rapidly becomes strong 
enough to affect 
the fluid motions, resulting in dynamo saturation and formation of the 
eventual magnetic configuration. The particular 
form of the saturation is still a disputed topic. 
However experience of galactic dynamos (see e.g.  Kleeorin et al. 
2000, 2002, 2003 where a sequence of more and more complicated models of dynamo saturation is used,  giving 
very similar saturated magnetic field configurations as in the simplest case) 
suggests that a reasonable 
approximation to the final configuration can be obtained using a
simple expression of the form

\begin{equation}
\alpha = \frac{\alpha_0}{1 + \xi B^2/B_{\rm eq}^2} \, ,
\label{sat}
\end{equation}
where $\alpha_0$ is the unsaturated value and $\xi$ is a numerical factor of order unity. Below we use this 
approach taking $\xi =1$ and bearing in mind the approximate nature of the approach. Note that changing the value of 
$\xi$ merely rescales magnetic
field in the saturated solutions.

\section{ alpha-quenched model}

\subsection{Preliminaries}
\label{prel}

Following the idea of our paper, we are interested 
in relatively simple models for dynamo driven magnetic 
configurations. Experience from models for galactic dynamos suggests 
that a substantial simplification of the general mean-field 
dynamo equations is also possible for accretion discs. 
Indeed, for the basic model, we expect excitation of 
a solution symmetric with respect to the disc equatorial plane, i.e. a configuration with quadrupole symmetry. 
Galactic dynamo studies suggest a method to reduce the mean-field equations for such a configuration to relatively 
simple forms, which give  mean or integral values taken perpendicular to the disc of magnetic field components 
parallel to the disc (this is known as the no-$z$ approximation: Subramanian \& Mestel 1993, Moss 1995). On the other hand, 
the dynamo drivers have axisymmetric 
distributions and it looks plausible that the dynamo driven magnetic 
may also be axisymmetric.

We start our analysis from some trial solutions constructed using this approximation. Solutions were 
found assuming 
axisymmetry (i.e. a 1D problem. The marginal case occurs for $\sigma_v\sim 1.54$, i.e. $D\sim 3.8$.  
Then a $\phi$-dependent case in 2D was run (where $\phi$ is the azimuthal coordinate).
Although the latter 
solutions were not followed to saturation, there was a very strong indication that saturated solutions would be 
axisymmetric (in
accordance with naive expectations). Thus, through the remainder of this paper
axisymmetry is assumed.
However, we cannot exclude the possibility that for very supercritical solutions
axisymmetry may be lost.

\subsection{ Basic 1D model}
\label{1Dres}

The inner computational boundary
is placed at the inner disc boundary $r=r_{\rm in}=0.01$.
We use 4001 uniformly spaced grid points which seems to be more than sufficient.
For this approximation, the inner boundary condition is rather arbitrary, and 
we simply extrapolate $B_r$
from the next two points,
and determine $B_\phi(\rm in)$ by requiring the ratio
$B^2({\rm in})/B^2_{\rm eq}({\rm in})$ to have the same value as at the next point
(the global solution is rather insensitive to this b.c.).

\begin{figure*}
\includegraphics[width=0.95\columnwidth]{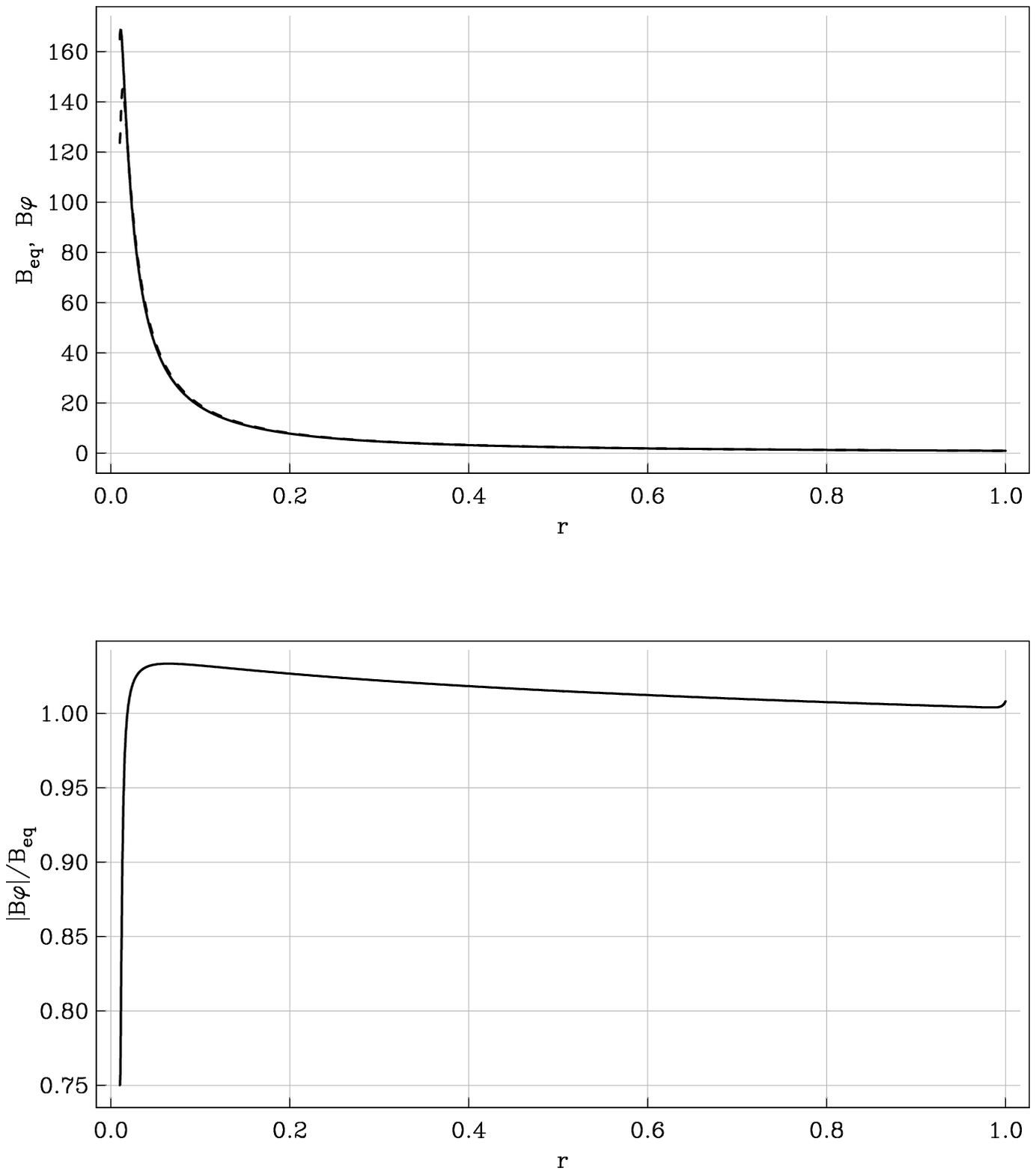} 
\includegraphics[width=0.95\columnwidth]{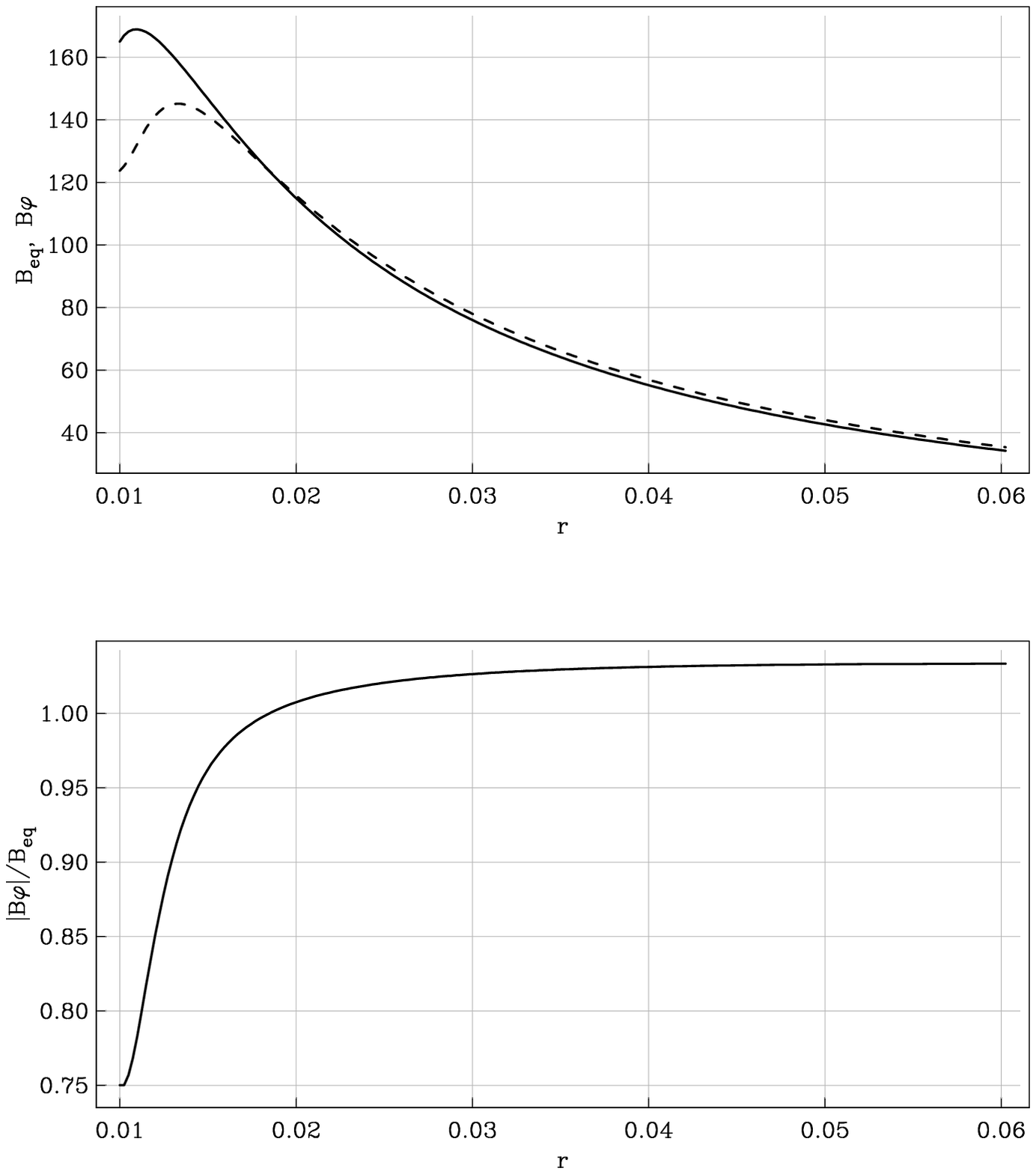} 
\caption{Global solution (left) and a magnified view of the solution near $r=r_{in}$ (right), for the slightly 
supercritical parameters 
$R_\alpha=0.6$, $R_\omega=15$, $R_m=0.146$. In upper panels solid line is for $B_{\rm eq}$  
and dashed line stands for $B_\phi$.}
\label{stand2} 
\end{figure*}

The final (saturated) magnetic configuration is shown in Fig.~1 for the basic model. We can experiment with the model,
keeping the main parameter $D$ constant. 
 
The steep rise in field towards $r=r_{in}$ appears robust (e.g. to change of b.c., and to
extension of the computational domain to $r<r_{in}$ as well as to enlargement of the infall velocity).
Typically, $B_\phi$ exceeds very slightly the equipartition value throughout the disc. 
The solution is rather "flat" w.r.t. $r$ over most of the range, presumably a consequence 
partly at least of the constant local dynamo number and the relatively small 
dependence of $B_{\rm eq}$ on $r$ in the outer regions of the disc.

We then experimented with enhancing the infall speed in the inner part of the disc, to
simulate the fall of material onto the central object. For example,
we enhance  the radial velocity in the innermost disc by putting
$u_{r*}(r)=u_r(r)*(0.10/r)^4$ in $r<0.10$, where $u_r(r)$ is that given by the 
standard solution,
and $u_{r*}(r)$ is the velocity used. There are just modest changes to the inner solution. In particular the
local maxima of $B_r, B_\phi$ are removed.
This suggests that very strong enhancements to $u_r$ are required for significant changes.

\section{The 2D solutions}
\label{2Dsoln}

Of course, the above 1D approach gives only some preliminary orientation, 
and its extension to two dimensions seems natural. In 
particular, it enables consideration of magnetic configurations that
are not symmetric with respect to the disc plane (i.e. not quadrupole-like)
and in particular mixed parity solutions become accessible. Also
the  $z$-dependence of the dynamo driven magnetic field can be described.

\subsection{Extension of the disc model to two dimensions}

The canonical accretion disc model described in Sect.~\ref{basic} has no $z$-dependence. To generate a model suitable for studying (axisymmetric) dynamo action
in two spatial dimensions the basic model must be 
extended (necessarily in a rather arbitrary fashion) to depend on the vertical coordinate $z$.
In the following the subscript d refers to values in the 1D model of 
Sect.~\ref{basic}. $\eta_{\rm m}$ is defined to be the asymptotic value of  the turbulent diffusivity $\eta$ 
far from
the disc. This is a commonly made assumption in embedded disc models for
galaxies, and is made to approximate a vacuum field at large distances. Taking into account that vacuum 
formally corresponds to infinite $\eta$ we assume that $\eta_m \gg \eta_d$. This important point is discussed 
in a galactic context by Sokoloff \& Shukurov  (1990): however other opinions 
have been presented in the literature.
Rather arbitrarily we  take $\eta_{\rm m}=5\eta_d$.

The $\alpha$ coefficient is written as

\begin{eqnarray}
\alpha(r,z)&=&\alpha_{\rm d}, |z|\le h_{\rm d}(r);\\
\alpha(r,z)&=&\alpha_{\rm d}\exp\left(-\frac{|z|-h_{\rm d}(r)}{2.5h_{\rm d}}\right)\frac{|z|}{z}, |h|>h_{\rm d}.
\label{alpdef1}
\end{eqnarray}
Similarly
\begin{eqnarray}
\eta(r,z)&=&\eta_{\rm d}(r), |z|\le 2.5h_{\rm d}(r);\\
\eta(r,z)&=&\eta_{\rm d}(r)(\eta_{\rm m}-\eta_{\rm d}(\left(r)\tanh(|z|-2.5h_{\rm d}(r)\right),\\ 
\nonumber && |z|>2.5h_{\rm d}(r).
\end{eqnarray}

\begin{eqnarray}
u_{\rm r}(r,z)&=&V_{\rm in}(r), |z|\le h_{\rm d}(r);\\
u_{\rm r}(r,z)&=&V_{\rm in}(r)\exp\left(-\frac{|z|-h_{\rm d}(r)}{2.5h_{\rm d}(r)}\right), |z|> 2.5h{\rm d}(r).
\end{eqnarray}

For the bulk of the computations $\Omega(r,z)=\Omega_{\rm d}(r)$.
In two trial cases $\Omega(r,z)$  decreased with distance from the disc plane
in a rather arbitrary manner -- see Sect.~\ref{basicmodel}.
Subscript d denotes a property of the 1D disc model.

\subsection{Code}

The code used was written in cylindrical $(r,z)$ geometry, based loosely 
on that of Moss \& Shukurov (2004).  Given the assumed axisymmetry,
the poloidal component of the magnetic field can be written as 
${\bf B}_{\rm pol}=\nabla\times A_\phi$, where $\phi$ is the azimuthal 
coordinate, and the problem then reduces to solving the evolution
equations for $A_\phi$ and $B_\phi$. The principle numerical difficulties
to be overcome are the extremely small disc thickness near
the inner boundary at $r = r_{\rm in} = 0.01$ and the strong gradient of 
angular velocity near this inner boundary.
The second difficulty was addressed quite satisfactorily by using $\log r$ as
the radial coordinate, with $n_{\rm r} =$ 151 or 301 points distributed 
uniformly between $\log(r_{\rm in})$ and $\log(1)$.
The computational domain included the disc plane $z=0$,
and computations were performed either in $0\le z\le z_{\rm m}$,
or in
$-z_{\rm m} \le z \le z_{\rm m}$. This made using a logarithmic vertical coordinate problematic, and it was decided 
to use the "brute force" method of taking
a closely spaced, uniform, vertical grid with $n_{\rm z}$ large enough to
resolve satisfactorily the inner regions of the disc.
With $z_{\rm m}=0.2$, taking $n_{\rm z}=8001$ was found to be adequate 
for calculations over the half-range, and $n_{\rm z}=16001$ for the full range.
Trial integrations for a supercritical case with $f_{\rm v}=1$ (Sect.~\ref{basicmodel})
in the half-space $z\ge 0$ confirmed that doubling the radial resolution to 
$n_r=301$ grid points, or increasing $n_z$ to 16001, did
not significantly affect the results.

The small spatial steps mean that a Runga-Kutte type code was not viable, 
as it would require 
extremely small time steps. Thus, as in Moss \& Shukurov (2004) a Dufort-Frankel 
integrator was used; $\Delta \tau=5\times 10^{-5}$ was found to be satisfactory.

\subsection{Boundary conditions}
\label{bcs}

Boundary conditions are again necessarily somewhat arbitrary, but are chosen partly in
the light of what previous experience has shown to be plausible and to give
physically realistic results.

On the upper and lower boundaries $z=\pm z_{\rm m}$, 

$$\frac{\partial A_\phi}{\partial z}=\frac{\partial B_\phi}{\partial z}=0.$$
When the equations are solved in the half space $0\le z \le z_{\rm}$ 
the boundary conditions on $z=0$ are determined by the imposed symmetry.
On $r=1$, 

$$\frac{\partial A_\phi}{\partial r}=\frac{\partial B_\phi}{\partial r}=0.$$
On $r=r_{\rm in}$, $$A_\phi=B_\phi=0,$$ so that $B_{\rm r}=0$ there.
In practice, the solutions show that field is very small at the boundaries
$z=\pm z_{\rm m}$ and $r=1$ so the exact boundary conditions  might be expected 
to have little effect on the solution.

To verify this, an alternative condition of extrapolating $A_\phi$ at
$r=r_{\rm in}$ from the values at the next two radial points,
and requiring that the ratio $|B_\phi|/B_{\rm eq}$ at $r=r_{\rm in}$ be equal
to that at the next radial point, was used in one case. Except very near
$r=r_{\rm in}$, there was little change in the solution.

\subsection{Procedure}

If we try to compare 1D and 2D solutions, some tuning of the dynamo numbers
 is needed. Essentially this is because of the (rather arbitrary) extension of 
the 1D disc model into the vertical direction; the input of the 1D model here is 
via the disc half-thickness H, and exact comparison of the models is impossible.
See also R\"udiger et al. (1995).
We want freely to explore parameter space, initially to find models near the
excitation threshold and then more generally,
and so again use the numerical parameter $f_{\rm v}$ multiplying $R_\alpha, R_\omega$ and $R_m$ introduced 
previously.
$f_{\rm v}$ is related to $\sigma_v^{-1}$ (Sect.~\ref{basic}); we retain the notation $f_{\rm v}$ 
with  $f_{\rm v} = \sigma_v^{-1}$
to emphasize that we deal here with a more detailed model.
Note again that $f_{\rm v}$ multiplies each of the dynamo numbers.
We still consider $R_\alpha=0.6, R_\omega=15$, $R_m=0.1254$ to be canonical 
values (i.e. $f_{\rm v}=1$). We note again that $f_{\rm v}<1$ corresponds formally
in our model to the physically unrealistic case of supersonic turbulence. However,
as discussed earlier, we think it useful to explore such cases.

Based on the usual explanation of the $\alpha$-effect as 
being related to the mirror asymmetry of turbulence caused by the Coriolis
force in a stratified medium, we expect that $R_\alpha >0$.
However taking into account 
the uncertainties intrinsic to the physics, solutions with 
negative values of the dynamo number $R_\alpha$ are also discussed.

\subsection{$R_\alpha>0$}
\label{basicmodel}

\begin{figure*}
\includegraphics[width=0.95\columnwidth]{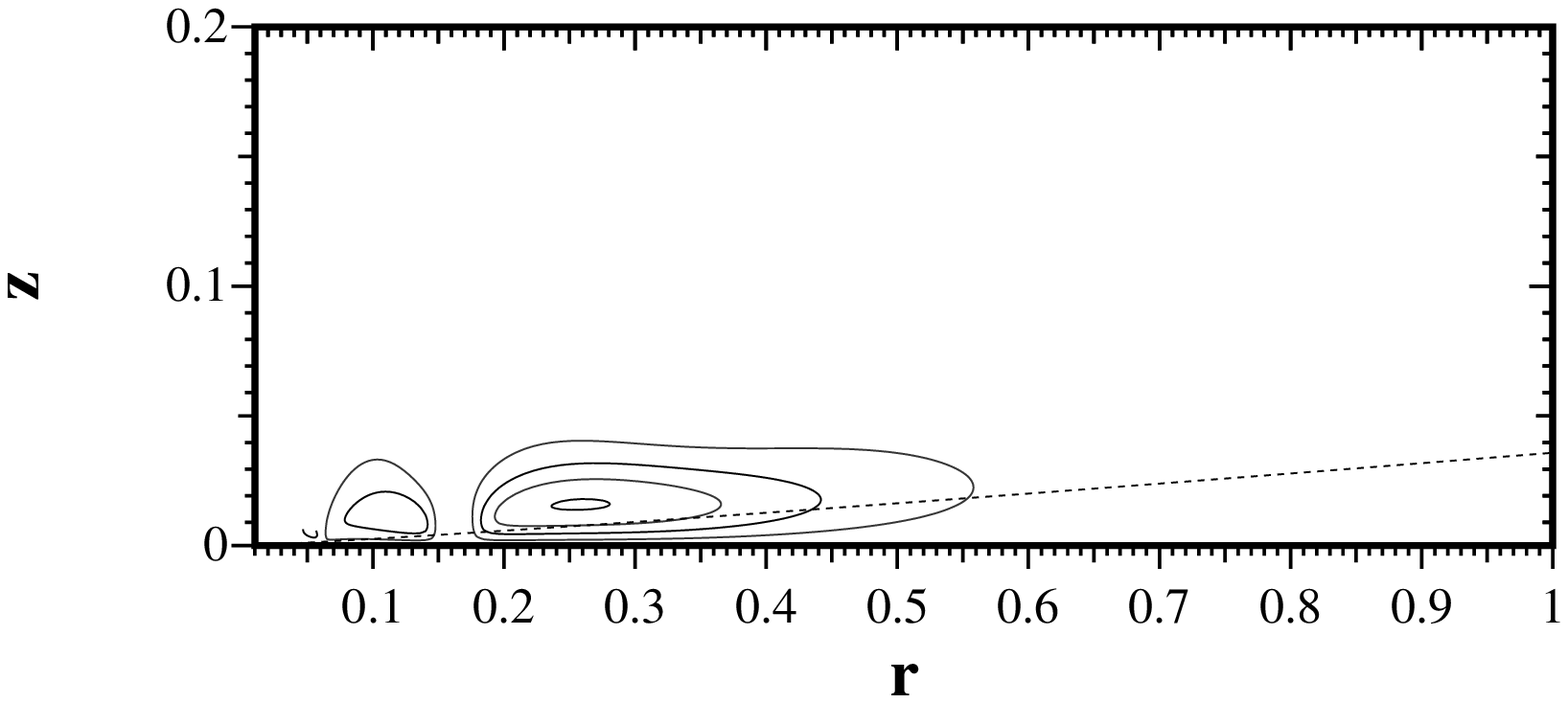} 
\includegraphics[width=0.95\columnwidth]{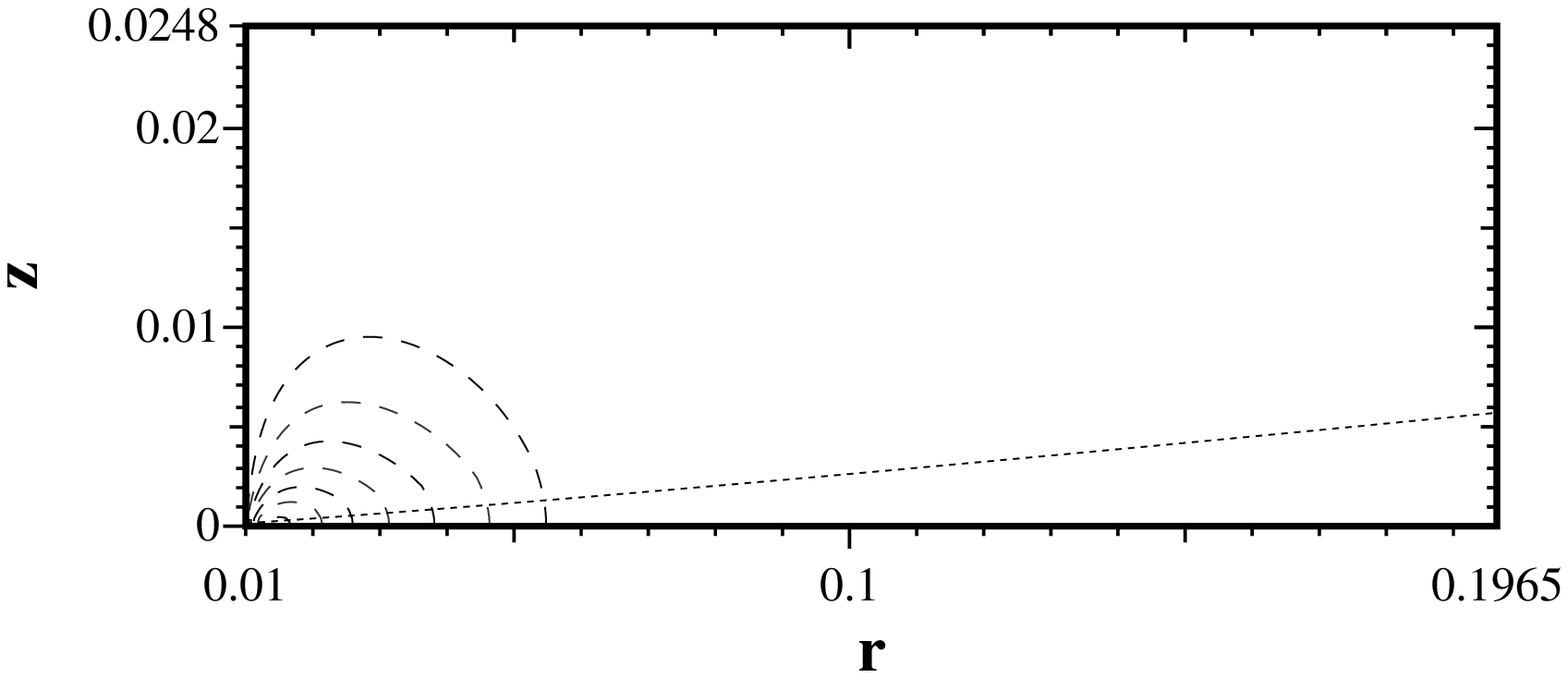} 
\caption{Global solution (left, contours of the poloidal field) and a magnified view of the solution 
near $r=r_{in}$ (right, contours of toroidal field), for the somewhat 
supercritical parameters in the 2D model ($f_{\rm v} = 0.3)$. Even parity, $P=+1$.
In this and subsequent figures, contours are equally spaced,
solid contours correspond to positive values, broken contours
represent negative values, and the dashed curve shows the disc boundary.
In the right hand panel the maximum toroidal field strength occurs
at $r\approx 0.015, z=0$ and is approximately $4 \times 10^5$\,G. The contour interval
is approximately $8 \times 10^4$\,G. In all the models the poloidal field is weaker: 
here the mean global ratio of poloidal to toroidal field strengths is about $10^{-2}$. 
In this and subsequent Figures, the field strength beyond the
outermost contour is smaller than that at the outermost contour. 
}
\label{stand} 
\end{figure*}

For the basic model we found that  $f_{\rm v}=0.3$ gives  a solution somewhat above the excitation threshold. This solution 
(Fig.~\ref{stand}) is steady, symmetric with respect to the central plane
of the disc (i.e. of quadrupolar symmetry) and generally resembles the
corresponding solution in the 1D model.
The radial extent of this solution is about  $[r]=0.05$, 
the maximum value of $|B_\phi|$  is below the equipartition  
value, ($|B_\phi|/B_{\rm eq} \la 0.4$, and $|B_\phi/B_r| >>1$  
(the ratio of global poloidal to toroidal field energies  $\sim10^{-4}$).

\begin{figure}
\includegraphics[width=0.95\columnwidth]{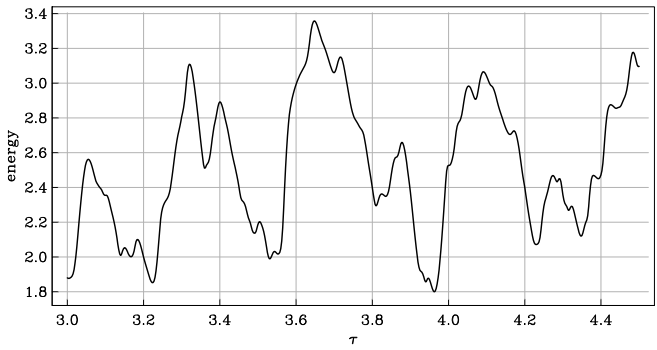}
\caption{Temporal behaviour of  the total global magnetic energy  
for
$R_\alpha=0.6$, $f_{\rm v}=1$. Even parity, $P=+1$.
(Energy is  normalized with the equipartition 
energy at the outer radius -- see Eq.~(\ref{Beq})).
Energy here and below is calculated by integration over the whole 
entire computaional volume (see Appendix for details)}

\label{Efveq1}
\end{figure}

\begin{figure}
\includegraphics[width=0.95\columnwidth]{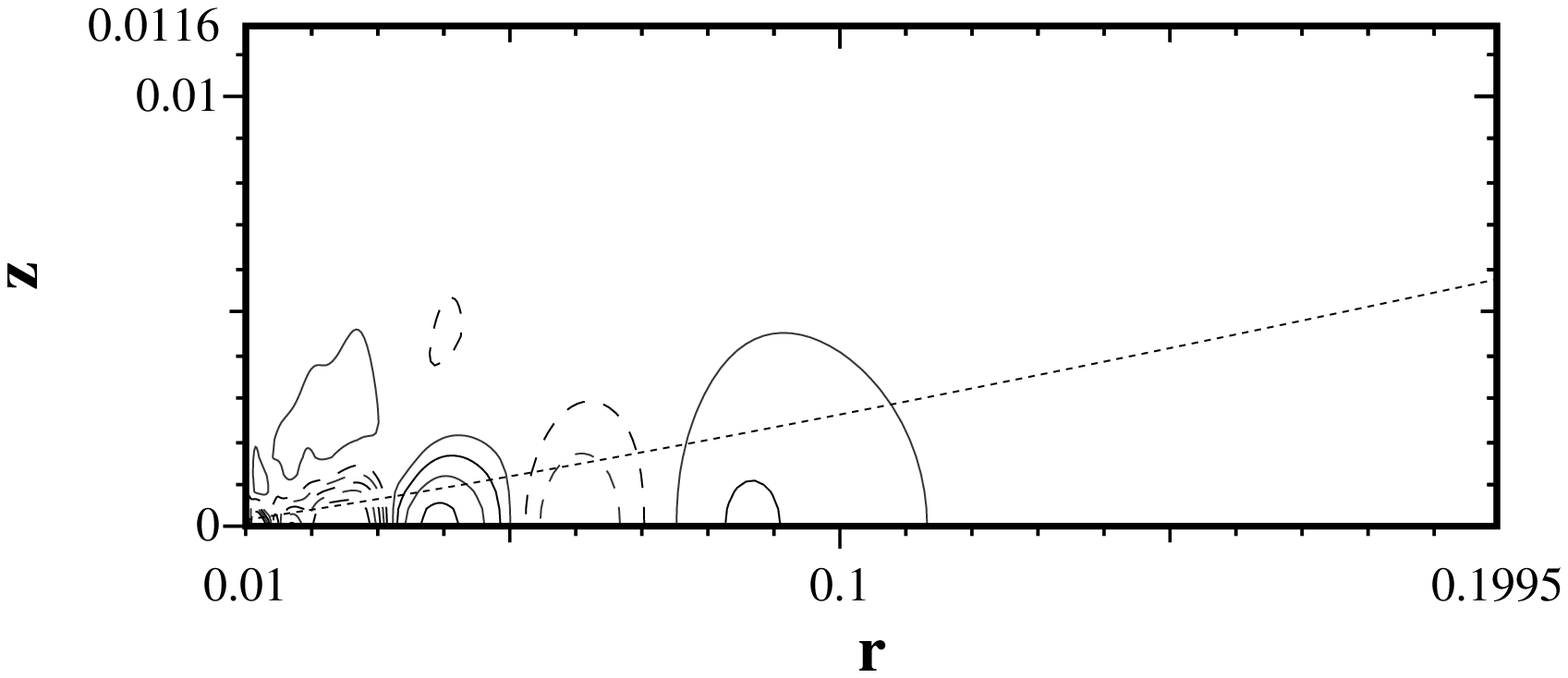}\\
\includegraphics[width=0.95\columnwidth]{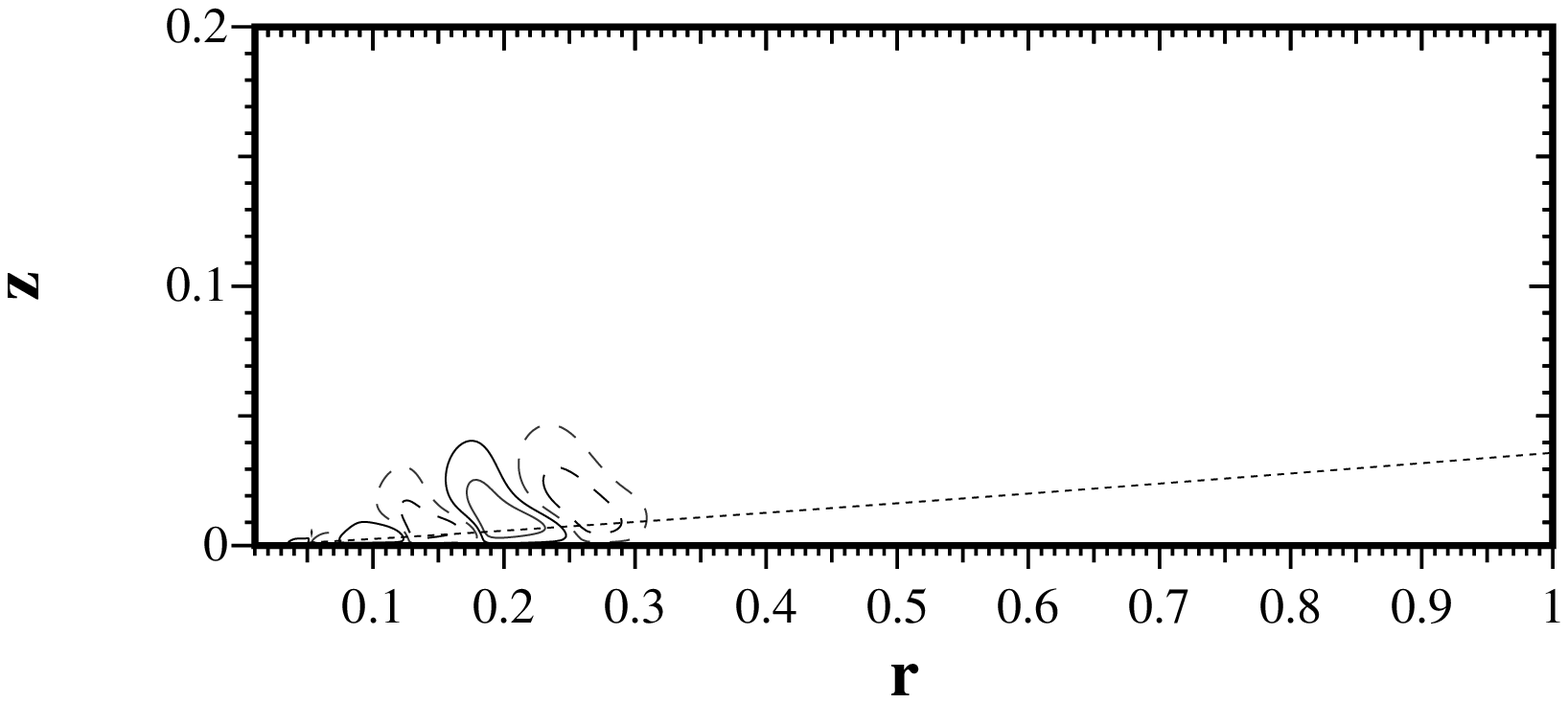}\\
\includegraphics[width=0.95\columnwidth]{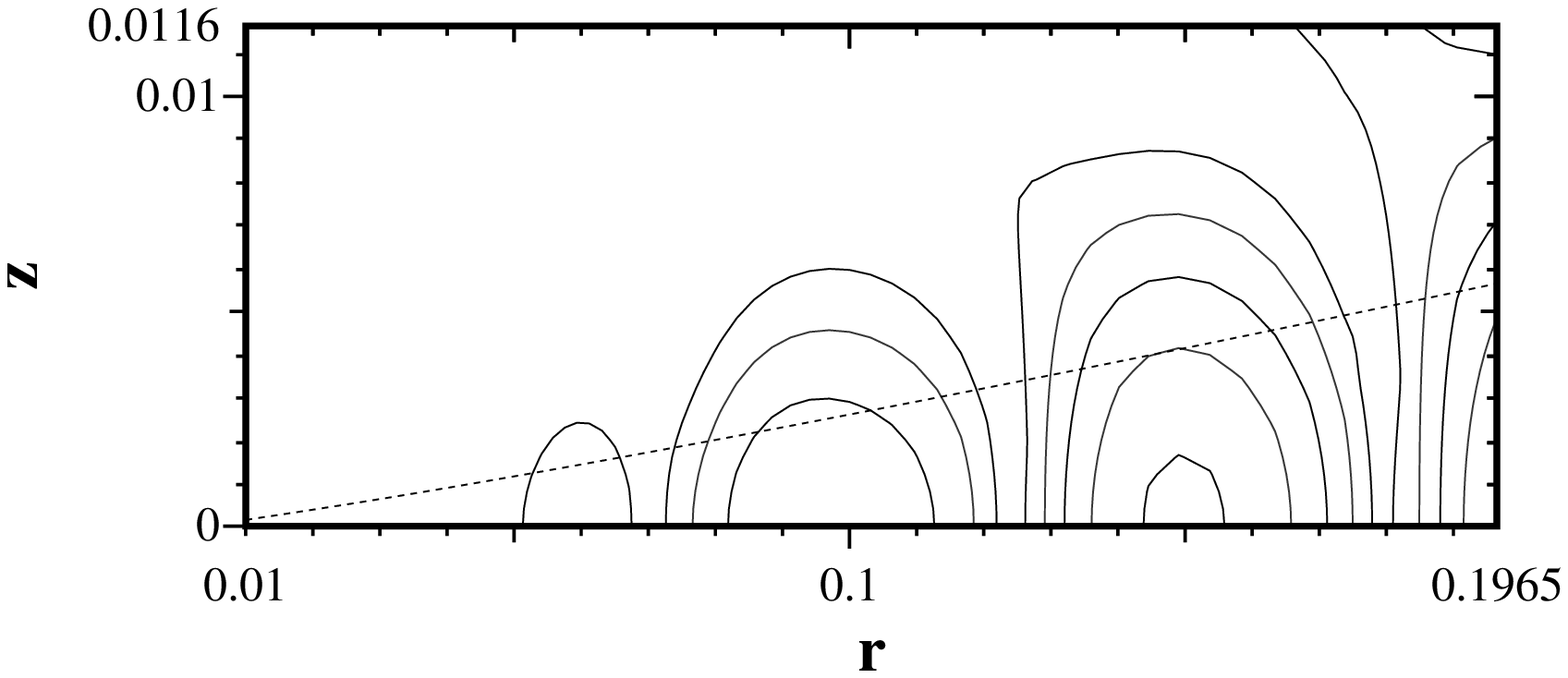}
\caption{Solution with $R_\alpha>0, f_{\rm v}=1$, even parity $P=+1$. The
contours lines of $B_\phi$ near the inner boundary (top panel),
the poloidal field lines in the complete computational space (middle)
and contours of the ratio $|B_\phi|/B_{\rm{eq}}$ (bottom).
In the top panel the maximum of the toroidal field strength is about
$2.4 \times 10^5$\,G, and the contour spacing is approximately $4 \times 10^4$\,G.
In the lower panel the maximum of $|{\bf B}|/B_{\rm eq}$ is about 0.6 at $r\approx 0.15, z=0$, with contour spacing 0.12.
 In this and subsequent Figures, contours are equally spaced.}
\label{Bfveq1}
\end{figure}

For more supercritical parameters such a steady homogeneous state
may be unstable, and indeed 
the temporal behaviour of saturated magnetic configurations becomes more complicated as dynamo action becomes stronger. 
With $f_{\rm v}$=1, the solution saturates with irregular temporal oscillations
in the total magnetic energy integrated over the computational space -- see Fig.~\ref{Efveq1}. 
The oscillations are approximately vacillatory, with no significant migration. 
The magnetic field is strongly concentrated to the disc plane in the inner disc,
the toroidal field more so than the poloidal.
Fig. ~\ref{Bfveq1} shows the contours of toroidal field and the poloidal
field lines. 
The maximum local ratio of field energy to
equipartition energy is approximately $0.6$. 
The lower panel of Fig.~\ref{Bfveq1} shows the contours of the ratio $|B_\phi|/B_{\rm eq}$.  If the computation is 
begun
with a seed field of odd parity and $f_{\rm v}=1$, there is no dynamo
action and the field rapidly decays. 
We define parity conventionally as $P= (E_{\rm even} - E_{\rm odd})/(E_{\rm even}+E_{\rm odd})$, 
where $E_{\rm odd}$ is the energy of the part of the field that is even 
with respect to the central plane of the disc,  and $E_{\rm odd}$ is 
energy of the odd part (more details are given in the Appendix). 

Given that $f_{\rm v}=0.3$ corresponds approximately to slightly supercritical 
excitation in this 2D case, this corresponds to 
$D \sim 1$ in the 1D model (see Sect.~\ref{basic}).
This can be compared with the value $D\la 4$ of Sect.~\ref{prel}.
There is a clear distinction between accretion disc dynamos and 
galactic dynamos, that is related presumably to the differing 
rotation curves, and maybe also to the disc geometry.
There is also a divergence between the 1D and 2D solutions; as the dynamo number increases, cell structures
are present in the 2D case, but not in the 1D. We speculate that these differences
may have a number of causes. For example, the rotation law is Keplerian vs. more-or-less flat for galaxies. 
Secondly, there is now significant 2D structure in the model, and in the
more supercritical cases the $z$-scale in the disc may impose a similar
horizontal scale (somewhat analogous to the cell structure seen in 
some spherical thin shell dynamo models, e.g. Moss \& Tuominen 1990).
It remains somewhat unexpected that the no-$z$ approximation seems to
break down for these more supercritical cases. We can speculate that this is 
because previous comparisons of the no-$z$ approximation with 
other models have been
made in galactic contexts, and for the slightly supercritical cases
thought to be relevant for galaxies (e.g. Phillips 2001; Chamandy et al. 2014).

\begin{figure}
\includegraphics[width=0.95\columnwidth]{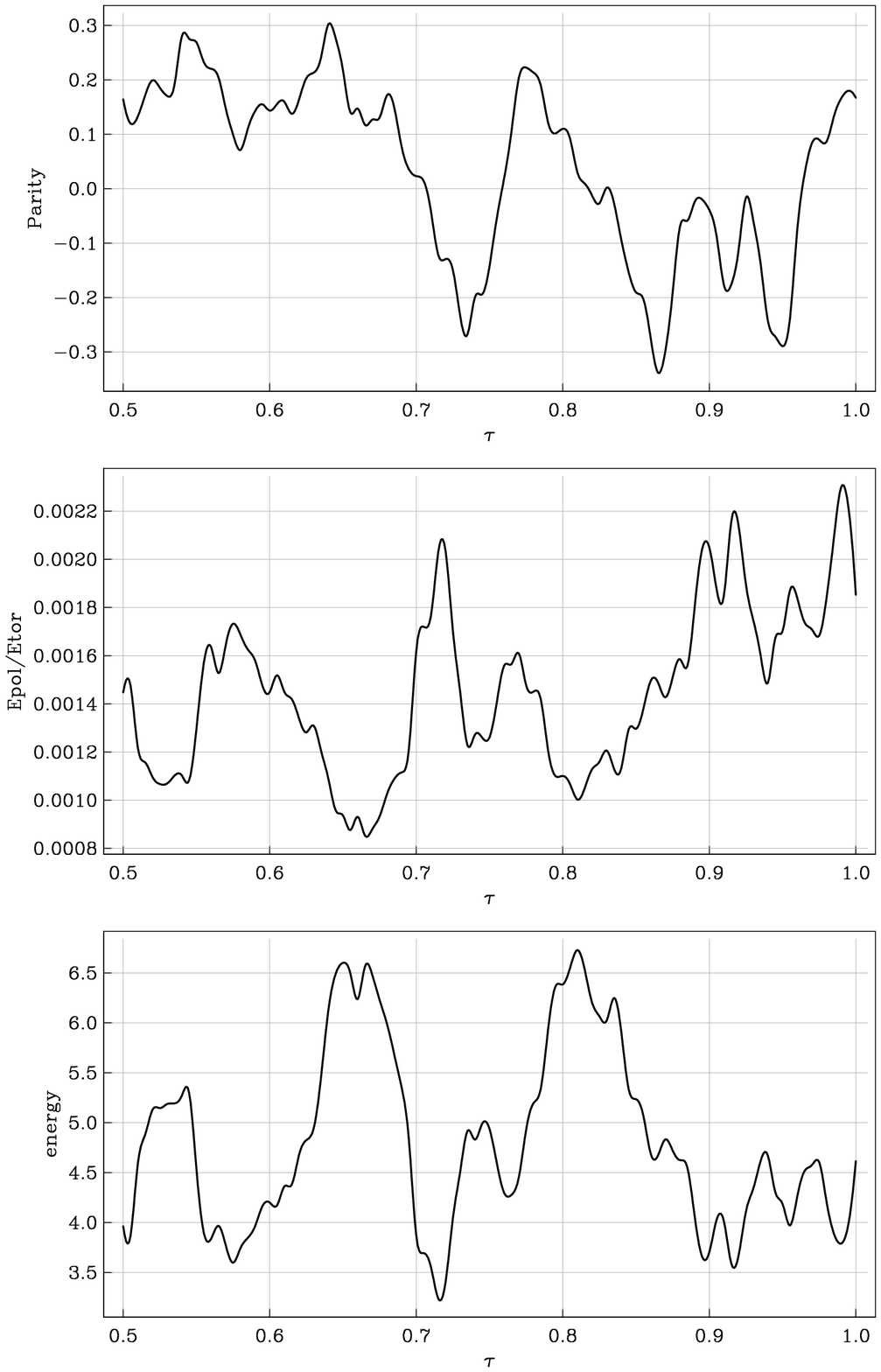}
\caption{Temporal behaviour of the total global magnetic energy for
$f_{\rm v}=2$;  solution with free parity. The upper panel shows the evolution of the parity $P$,
the middle panel the ratio of poloidal to toroidal energies, and the lower
panel the evolution of the total magnetic energy.}
\label{Efveq2}
\end{figure}

\begin{figure*}
\includegraphics[width=0.95\columnwidth]{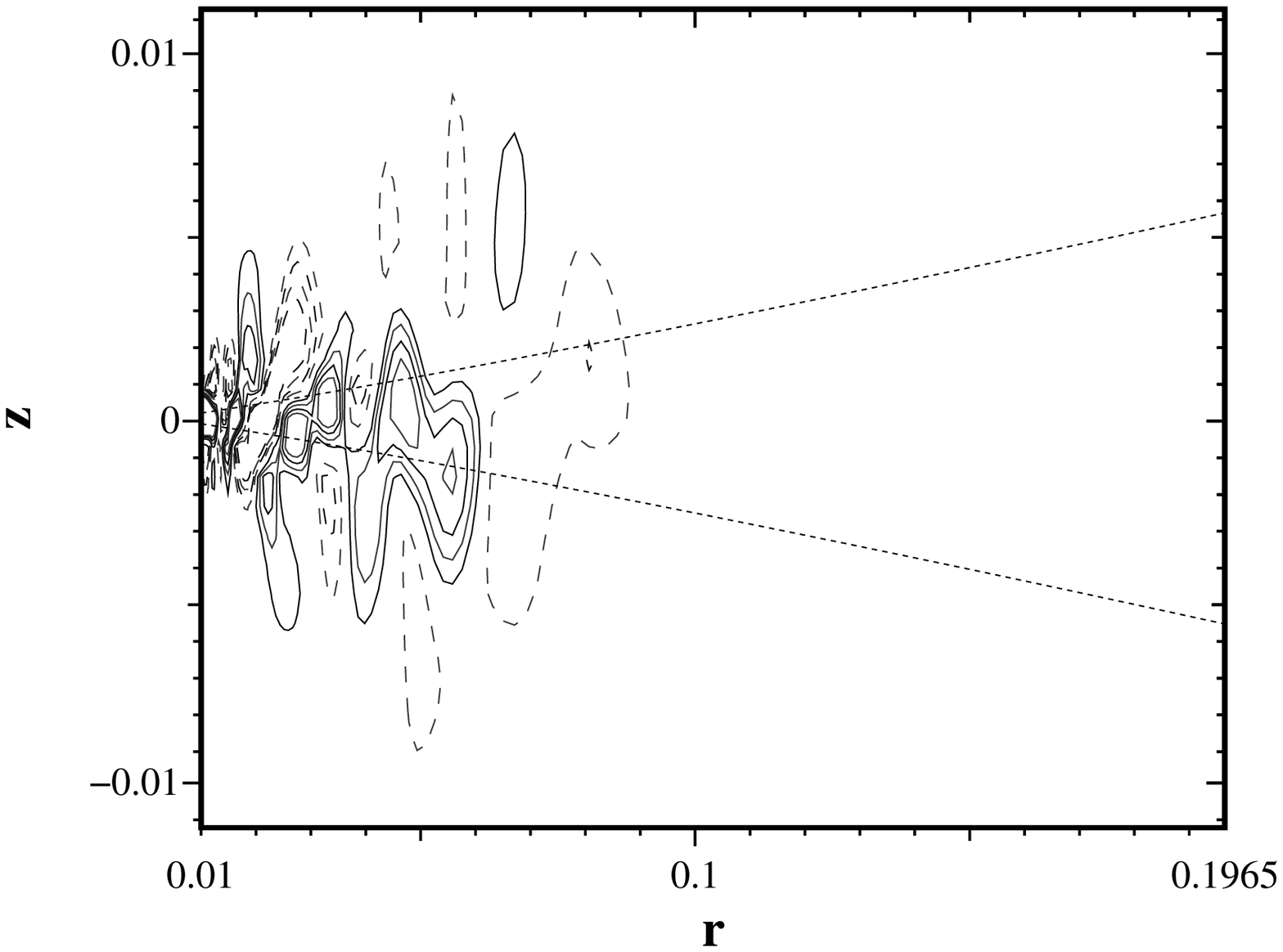}
\includegraphics[width=0.95\columnwidth]{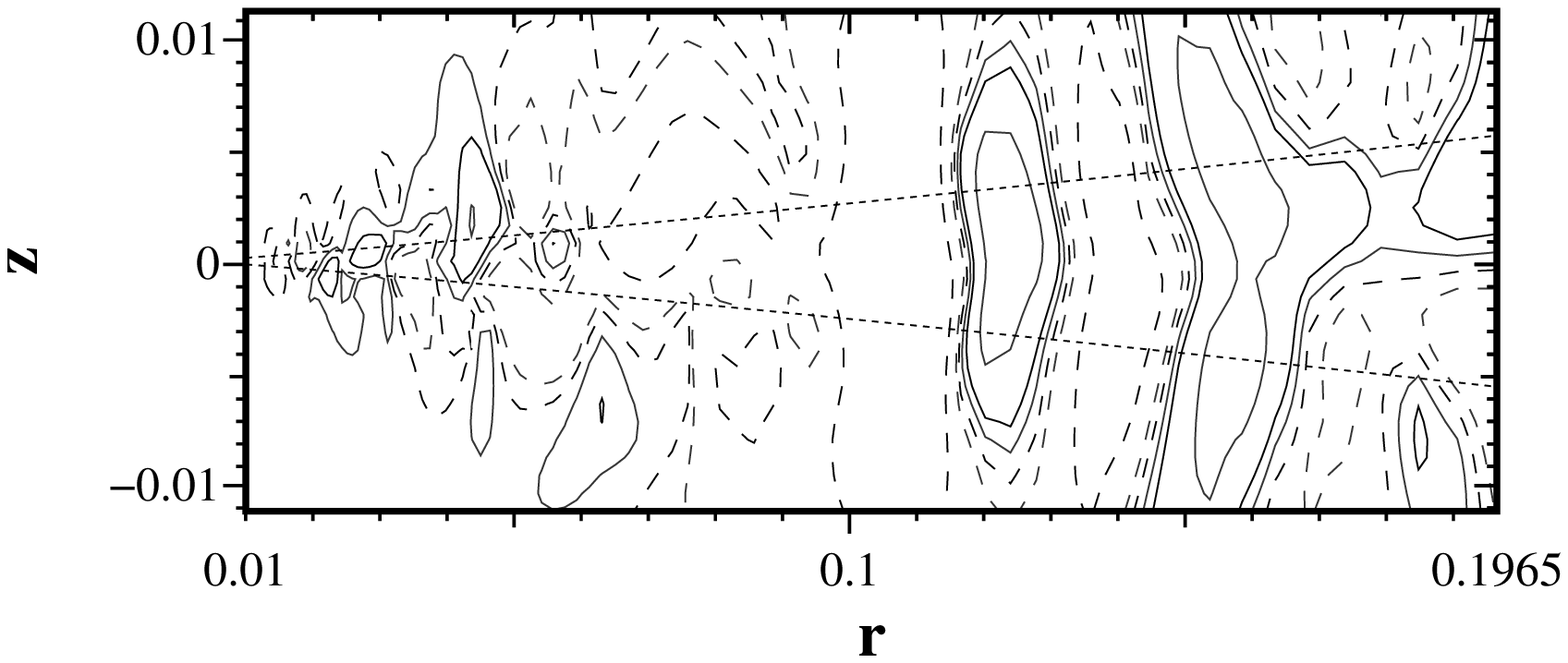}
\caption{2D solution with free parity and $R_\alpha>0, f_{\rm v}=2$; the computation
extends over $-z_{\rm m}\le z\le z_{\rm m}$. The
contour lines of $B_\phi$ near the inner boundary(left) and 
poloidal field lines (right) in the complete computational space.}
\label{Bfveq2}
\end{figure*}

Some computations were performed over the full space $-z_m\le z\le z_m$.
We confirmed that with $f_{\rm v}=1$, the solution discussed above
is recovered. For stronger excitation, the symmetry with respect to
the disc plane is lost.  
The temporal behaviour and spatial structure of the
solution with $f_{\rm v} = 2$ 
i.e. $D 
\approx 36$ in the 1D case) is shown in Figs.~\ref{Efveq2} and \ref{Bfveq2}
respectively.    

In this case, the magnetic configuration has acquired some of the 
characteristics of a symmetric dipole-like configuration. The toroidal magnetic 
field becomes concentrated near to the upper and lower boundaries of the 
accretion disc, instead of near  the central plane as 
found for  configurations with pure  quadrupolar symmetry. The radial magnetic 
field profile seems to be shifted 
more towards the inner boundary than previously.

In two cases with $f_{\rm v}=1$, $\Omega$ was allowed to vary with $z$. In the more 
extreme, $\Omega$ was independent of $z$ in $0\le z\le h(r)$, and decreased
with a scale height of $2.5 h(r)$ in $z>h$. The global energy of the
solution was significantly increased (presumably due to the increased dynamo 
efficiency caused by the steep gradient of $\Omega$ 
in the $z$-direction), and there was 
a little more radial structure. However the irregular temporal oscillations, 
and the general form of the solution, were very similar to those with
$\Omega$ independent of $z$. In the other case, $\Omega$ was independent of $z$
in $0\le z\le 0.1$, and then declined in $z>0.1$. Perhaps unsurprisingly,
this modification did not affect the solution strongly.

\subsection{$R_\alpha<0$}
\label{ralphaneg}

\begin{figure*}
\includegraphics[width=0.95\columnwidth]{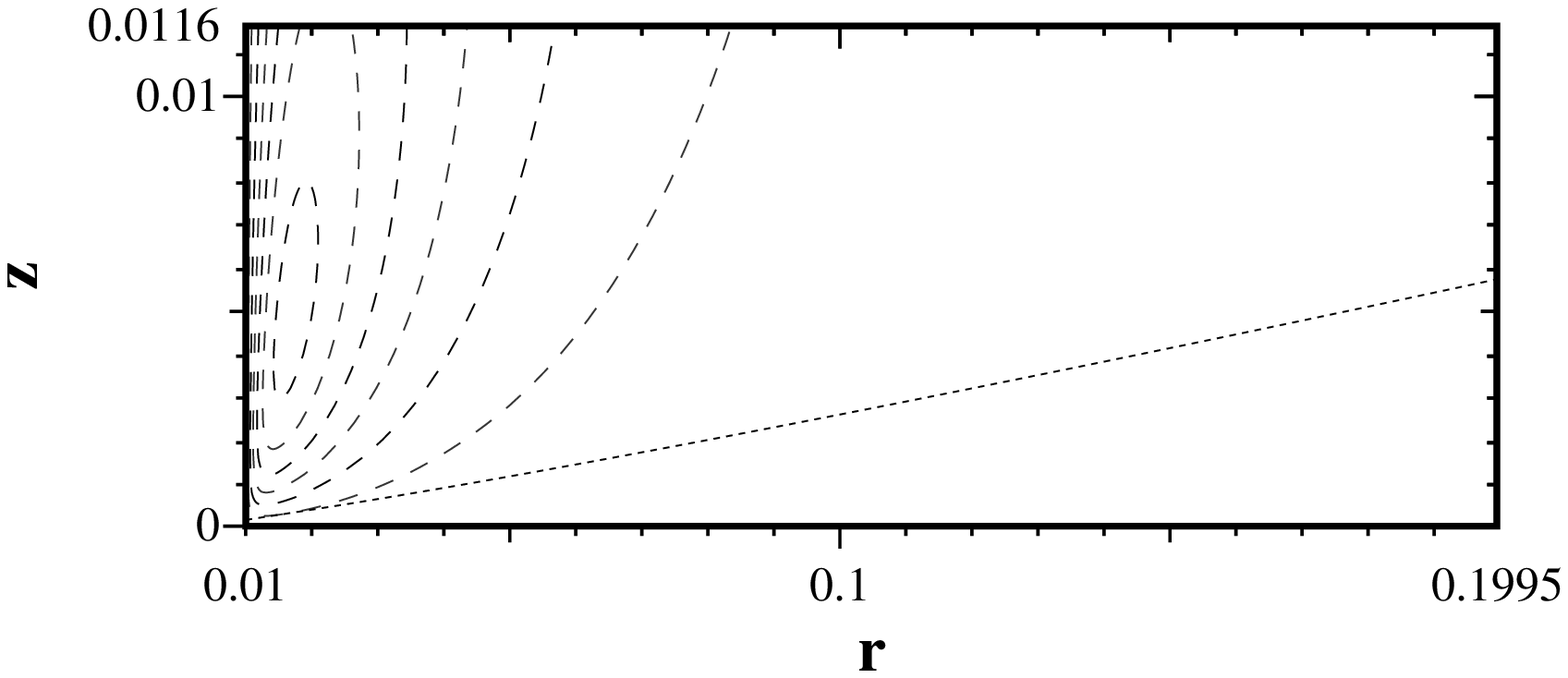}
\includegraphics[width=0.95\columnwidth]{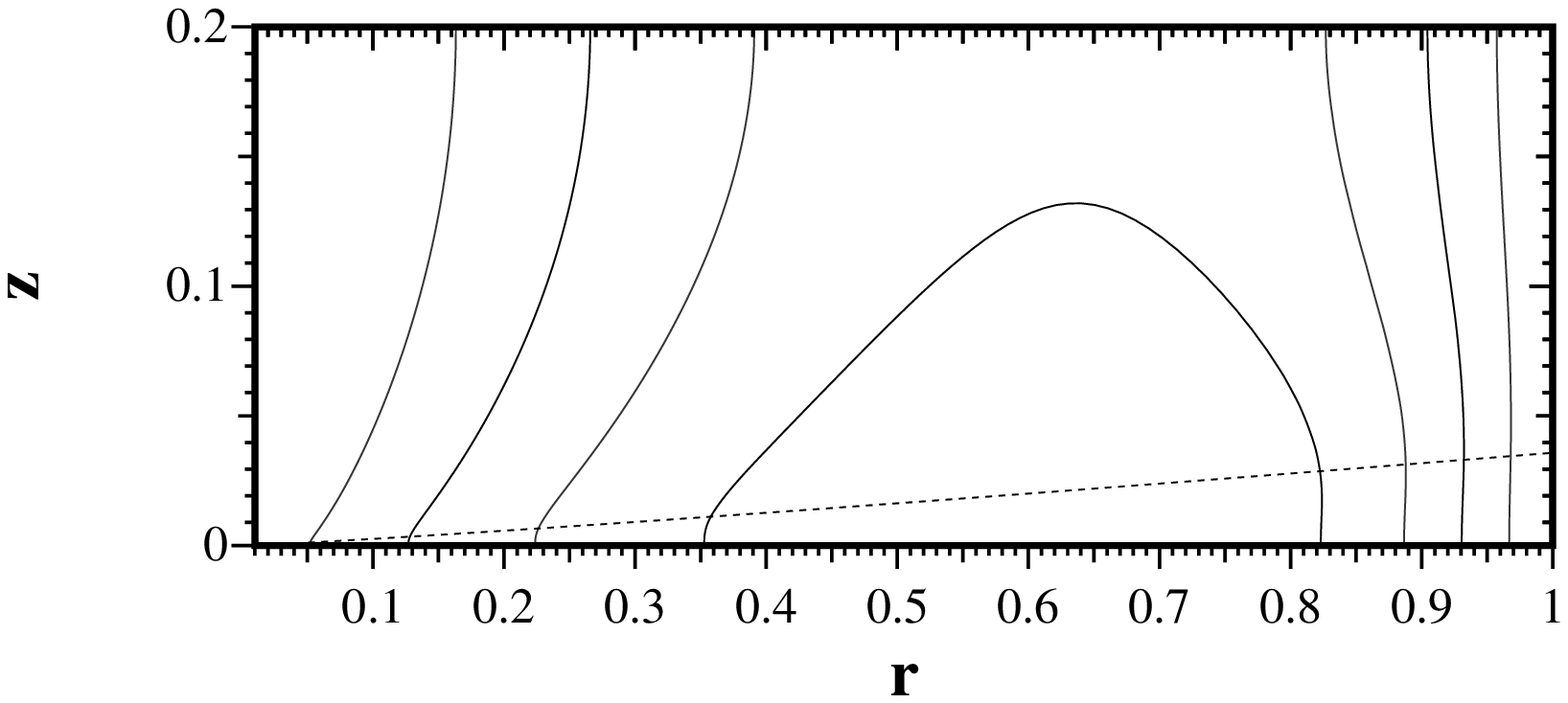}
\caption{Solution with $R_\alpha<0, f_{\rm v}=0.25$, odd parity $P=-1$. The
contour lines of $B_\phi$ near the inner boundary(left) and 
the poloidal field lines in the complete computational space (right).
The maximum value of $|B_\phi|$ is approximately $7.2 \times 10^5$\,G, and the contour spacing is $1.2 \times 10^5$\,G. 
The mean global ratio of poloidal to toroidal field strength is about $3 \times 10^{-3}$.}
\label{Ralpeq-1pt2}
\end{figure*}

Taking reference values as $R_\alpha=-0.6$, $R_\omega$ and $R_\alpha$ as above,
then
with $f_{\rm v}=0.25$ a finite amplitude odd parity solution is
found. The total energy approaches its steady asymptotic value smoothly.
The field structure is given in Fig.~\ref{Ralpeq-1pt2}.
Remarkably, the toroidal magnetic field is now concentrated above and below the disc rather than near the disc boundary. 
In our units scaled to approximately 400 G, the maximum value of $B_\phi$ is 
about 1000 whereas the poloidal field is very much weaker 
(the global ration of poloidal
to toroidal field energies is about $10^{-5}$).

\section{Models with Lorentz force}
\label{lorentz}
The strong fields found in the innermost disc suggest that the Lorentz force
may affect the rotation curve, with consequent effects on the dynamo action.
Thus we amended the code to include a representation of the Lorentz force 
from the large-scale magnetic field on the azimuthal motions.

\subsection{Amended model}
Following Moss \& Brooke (2000) we write
\begin{equation}
\partial v'/\partial t=\frac{\nabla\times{\bf B}\times {\bf B}}{\rho r}\cdot\hat\phi+{\rm viscous}\hspace{0.1cm} 
{\rm diffusion},\end{equation}
where $v'$ is the perturbation to the azimuthal velocity. Some smaller
terms involving advection of angular momentum by the accretion velocity
have been omitted.
We ignore the effect of the radial component of the Lorentz force; Rekowski et al. (2000)
 found it to be small, albeit in a disc model that did not extend to the small fractional radii considered here.
Note that the alpha-quenching non-linearity is also retained in the modelling.
In this part of the investigation we amended the basic rotation curve within radius 0.03, from the  Keplerian law 
to a linear law 
-- see the broken curves in Fig.~\ref{omeg_l39}.
The  immediate motivation for this softening  of  $\Omega$ is to
avoid computation difficulties, and to allow some progress to be made. 
With too rapid inward growth of angular velocity
saturated states could not be computed -- plausibly this is a consequence of the
restrictions on spatial resolution.
We note, however, that there is a physical motivation for the softening of
the angular velocity 
--
the angular velocity has to decrease to that of the white dwarf in the 
boundary layer between the 
accretion disc and the white dwarf. Typical theoretical boundary layer models 
have a radial extent comparable with the
radial width of the $\Omega$ softening suggested here (see, e.g. Hertfelder et al. 2013). 
Formally, deviations from the Keplerian law mean that our model is 
not now self-consistent at these radii, as the disc parameters
were computed using the Keplerian law. 
The simplification of ignoring this effect, however gives us the possibility 
of evaluating in a semi-quantitative 
way the size of the effect on the angular velocity
law that might be expected to arise from the back-reaction of the Lorentz force.
Of course this lack of self-consistency also applies to models in which
the angular velocity is significantly modified by the Lorentz force
(see figures below, such as Fig.\,\ref{omeg_l39}). 
It is clear, that the fully self-consistent accretion disc models 
with Lorentz force taken into account  need to be computed -- this is work for the future.

Solutions that converged to a (statistically) steady state could not
be found with $\alpha$ defined as in Eq.~(\ref{alpdef1}). We thus took
$\alpha$ to be non-zero in the disc only,

\begin{eqnarray}
\alpha(r,z)&=&\alpha_{\rm d}\sin(\pi z/h_{\rm d}(r)), |z|\le h_{\rm d}(r);\\
\alpha(r,z)&=&0, |h|>h_{\rm d}.
\label{alpdef2}
\end{eqnarray}
In other words, we have to exclude the exponential tail of the
 $\alpha$-distribution in the region surrounding the
disc and prescribe zero $\alpha$-effect in the surrounding space. 
Such tails were unimportant for  the
quasi-kinematic problem, however they prevent complete 
separation of dynamo action 
in the disc from processes in the surrounding 
space. As far as we know this point has not been reported in the galactic 
dynamo studies: arguably it is an outcome of the specific rotation law.
(The issue of a separation between
disc and halo dynamo actions has been considered in a galactic context by Moss \& Sokoloff, 2008).

Simultaneously, 
to obtain convergent solutions, it was found necessary to increase
the aspect ratio of the computational box; a ratio of 0.5 was found to be
safely adequate to obtain statistically steady states.
Now a vertical resolution $nk\ge 16001$ is required, and $nk=16001$ 
(and ni=151) was taken for most models. Some solutions were
recomputed with $nk=32001$, also with $ni$ increased to 301 and $nk=16001$.
Even so taking $r_{\rm in}=0.01$ as in the quasi-kinematical models
did not give satisfactory models, and most solutions used $r_{\rm in}=0.015$.
(We would have liked to compute solutions with $ni=301$, $nk=32001$, but this
resolution  was inaccessible with the facilities available.
Such an increase might allow taking $r_{\rm in}=0.01$.)

In the computations described below, the initial angular velocity associated
with the disc model was taken to be independent of $z$. We did experiment
with the possibility of a $z$-dependent initial angular velocity,
but could only find satisfactory solutions when the $z$-dependence was
very weak. Given that the important dynamo action occurs near to the disc,
where vertical variations in angular velocity can be expected to be small, 
we did not pursue this point.
Because of these difficulties, computations were 
now  performed  in the
region $z\ge 0$; previous experience strongly suggest that stable solutions are
strictly of even parity, at least for not very supercritical states,
so we confined our studies to even parity configurations.
This is also supported by the work of Rekowski et al. (2000).

Note that the change in definition of $\alpha$ from that in the earlier,
quasi-kinematic, models means that the the excitation conditions etc of the
two sets of models are not strictly comparable: for example, $\int \alpha dV$ over the
disc volume is now significantly reduced.

\subsection{Results: models with $R_\alpha>0$}
\label{res_even_lor}
We first investigated models with prescribed even parity. These arise when  $R_\alpha >0$, i.e. the case where 
right handed turbulent vortices dominate in the northern hemisphere of the disc. According to modern views 
of the origin of the $\alpha$-effect, this case is strongly preferred.

\begin{figure}
\includegraphics[width=0.95\columnwidth]{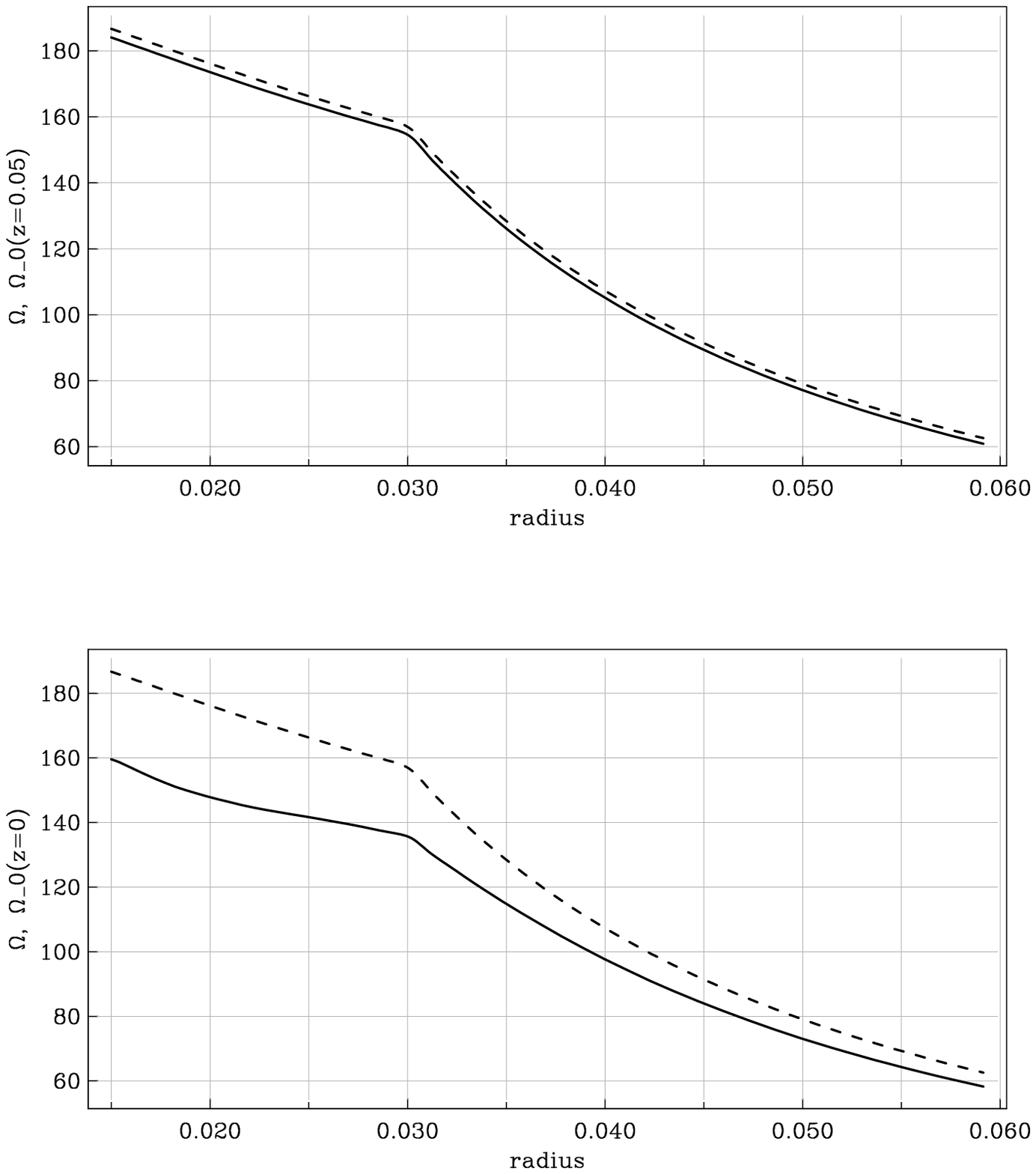}
\caption{Radial dependence of the modified angular velocity (solid)
and the initial angular velocity (broken)  at $z=0$ at $z=0.05$ 
(broken) at small radii for model l39 with $f_V=0.667$.  }
\label{omeg_l39}
\end{figure}

\begin{figure*}
\includegraphics[width=0.95\columnwidth]{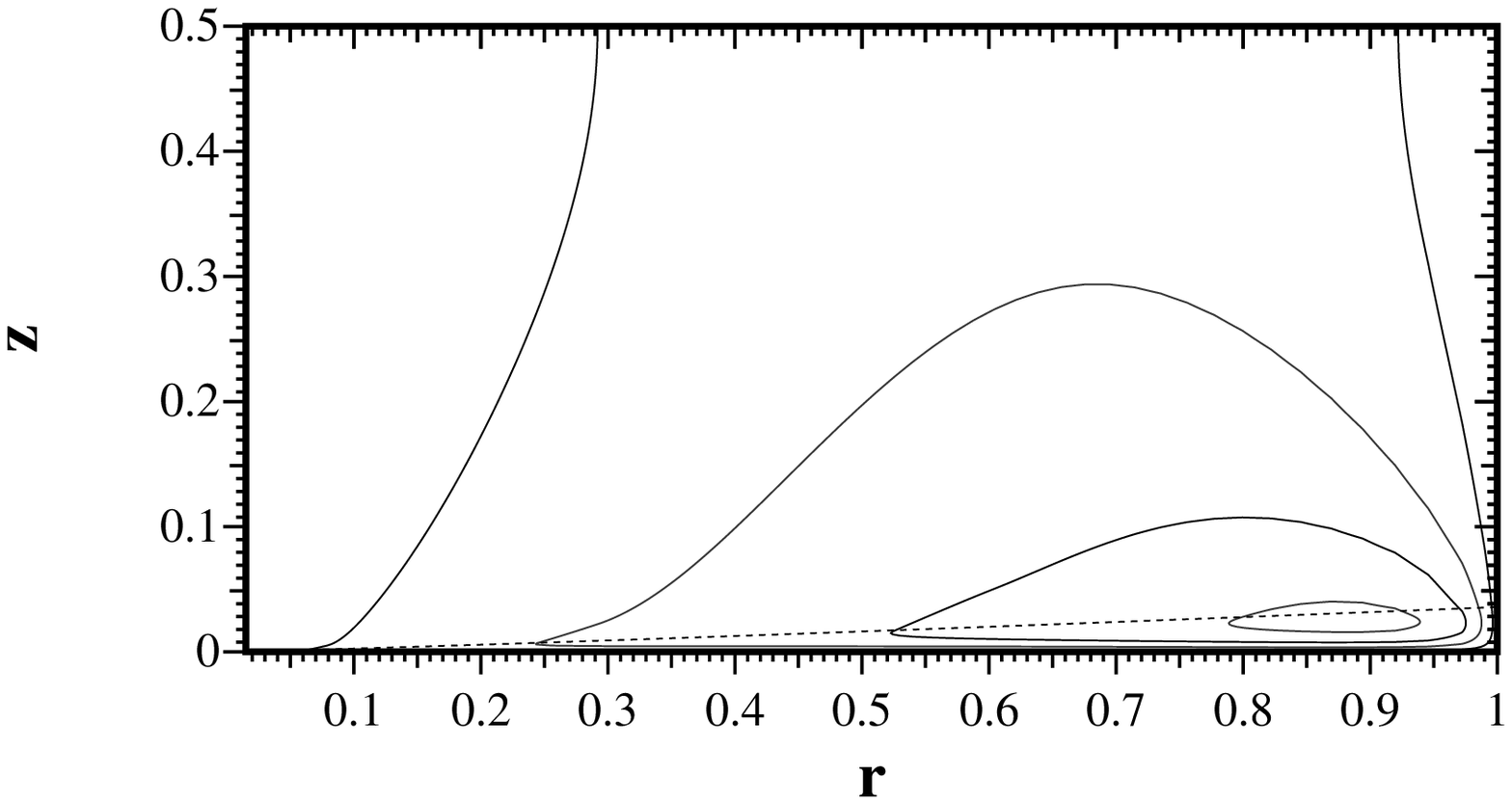} 
\includegraphics[width=0.95\columnwidth]{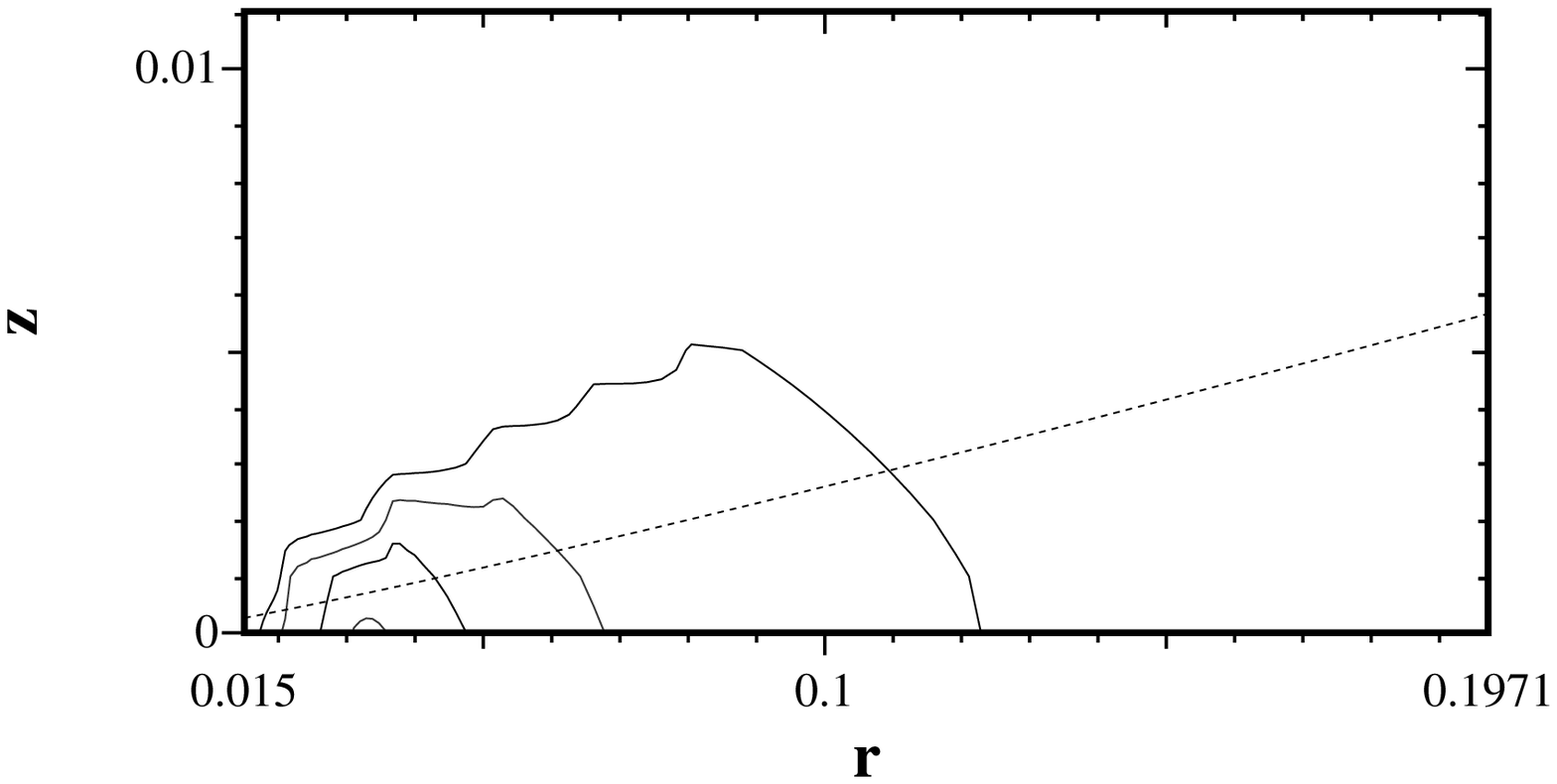} 
\caption{Global solution (left, contours of the poloidal field) and a magnified view of the solution 
near $r=r_{\rm in}$ (right, contours of toroidal field), for the slightly 
supercritical parameter for the even parity model with $f_{\rm v} = 0.667$ -- model l39. 
The dashed curve shows the disc boundary. Here and elsewhere wobbles in contours near the inner disc plane appear artefactual -- see the text.l}
\label{B_l39} 
\end{figure*}

Using the previous notation, with the slightly supercritical value 
$f_{\rm v}=0.667$ we obtain a strictly steady state,
with magnetic field more-or-less smoothly distributed over the disc (model l39).
Distortions of the rotation curve are minor, and 
restricted to the near disc region close to $r=r_{\rm in}$. Fig.~\ref{omeg_l39} shows
the variation with radius of the modified angular velocity and the  initial angular
velocity on $z=0$ and  $z=0.05$. Fig.~\ref{B_l39} shows the distribution
of the global poloidal and inner toroidal fields. The "wobbles" in the
toroidal field  distribution in the region above the disc are artefactual,
but experiments suggest that they do not appear to depend on the spatial 
resolution.

\begin{figure}
\includegraphics[width=0.95\columnwidth]{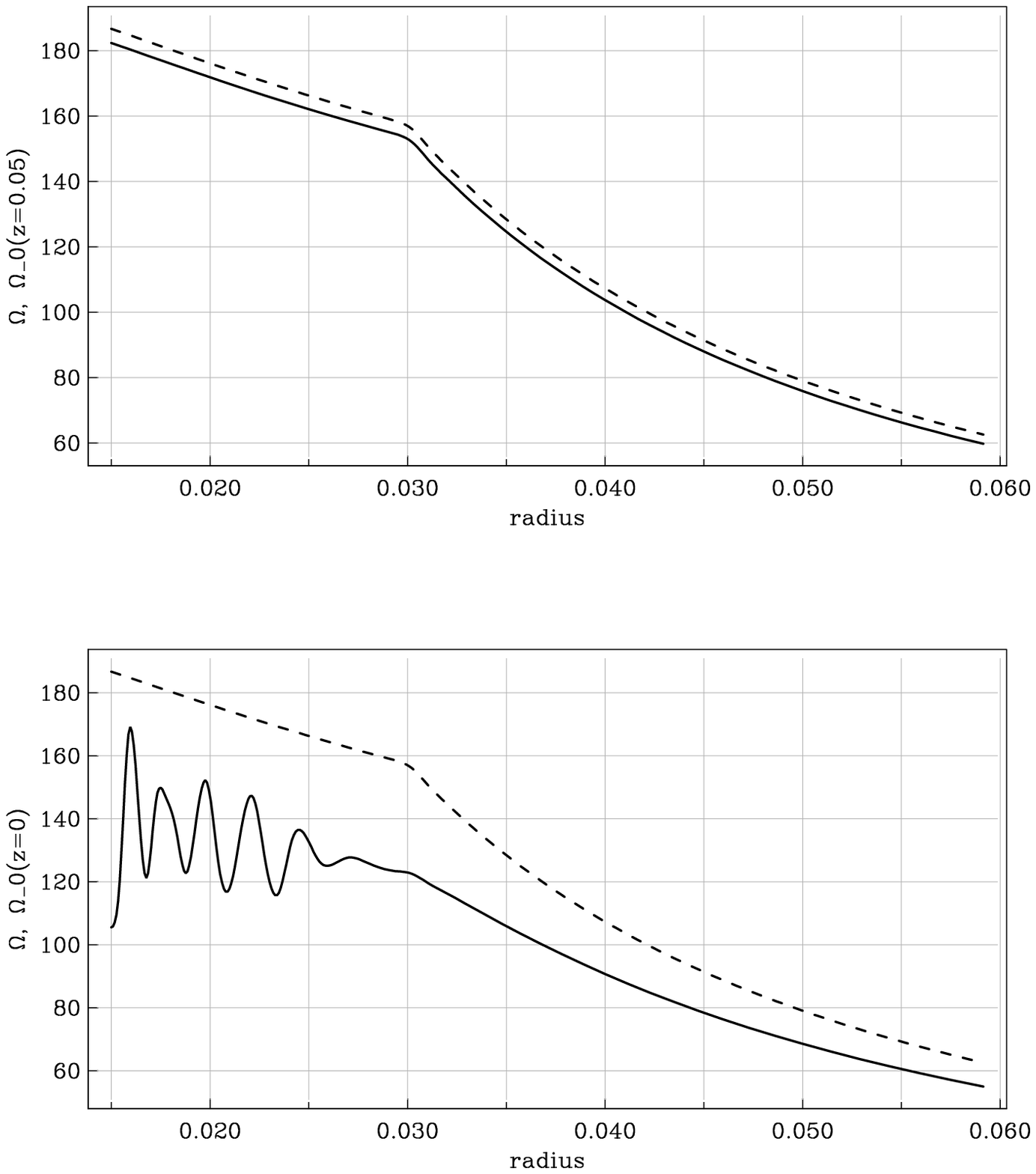}
\caption{Radial dependence of the modified angular velocity (solid)
and the initial angular velocity (broken)  at $z=0$ at $z=0.05$ 
(broken) at small radii for model l28.
  }
\label{omeg_rad_l28x}
\end{figure}

\begin{figure}
\includegraphics[width=0.95\columnwidth]{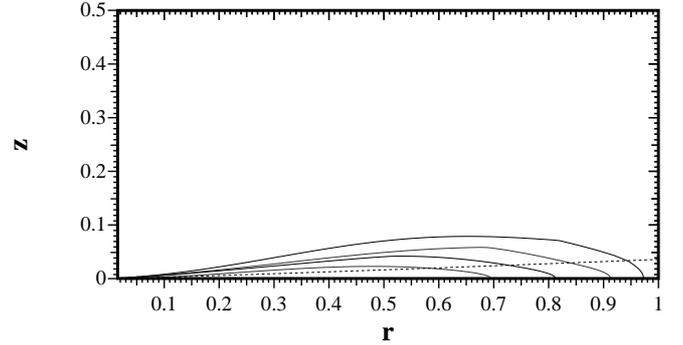}
\caption{Distribution of the ratio of total field strength to equipartition field strength for model l28. There are departures from the
smooth curves only at very small radii. The maximum is approximately 4.2, and contours are at values 1.0, 1.8, 2.6, 3.4.} 
\label{Beq_l28x}
\end{figure}

With $f_{\rm v}=1$ (model l28), the global energy develops very small 
irregular fluctuations 
$(\la 0.01\%)$ in the saturated state, and the angular velocity is noticeably
modulated by the Lorentz force -- see Fig.~\ref{omeg_rad_l28x}. This effect is again limited to small radii, 
with  a $z$-extent of a few multiples of the disc height. The energy fluctuations
arise from this region. $f_{\rm v}\approx 1$ marks a transition from steady states with field 
distributed smoothly with radius to fluctuating
states with toroidal field concentrated
strongly to the inner disc region. The distribution of the ratio of
total field strength to equipartition field strength may be more informative:
it is shown for the model with $f_{\rm v}=1$ in Fig.~\ref{Beq_l28x}. Apart from
the innermost regions, this distribution does not change greatly as 
$f_{\rm v}$ increases.

\begin{figure}
\includegraphics[width=0.95\columnwidth]{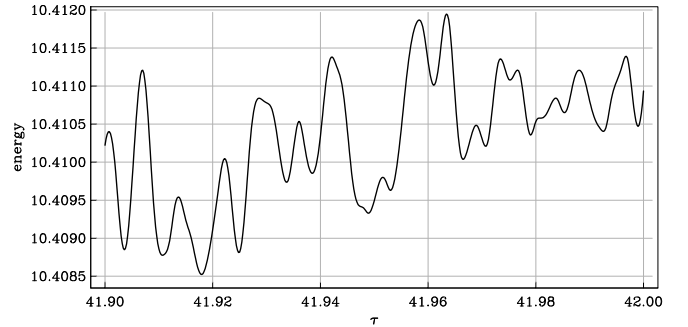}
\caption{Variation with time of global magnetic energy for model l36 with
$f_V=1.5$.}
\label{l36_E}
\end{figure}

With $f_{\rm v}=1.5$ (model l36), the global energy has quite irregular fluctuations
(Fig.~\ref{l36_E}), and the field continues to be concentrated in the inner disc
region.  

Attempts were made to obtain models with larger values of $f_{\rm v}$, but then greater spatial
resolution than available seems to be required to obtain reliable results.
As stated, the 'standard' case has resolution 151x16001 grid points.
Cases were also run with 301x16001 and 151x32001 points. There were some
differences, but these experiments strongly suggest that the trends seen
when moving
from $f_{\rm v}=1$ to $f_{\rm v}=1.5$ continue for larger values of $f_{\rm v}$: 
the field continues to be concentrated
near the inner disc, and this region is the source of the temporal fluctuations.

\subsection{Results: models with negative $R_\alpha$}
\label{res_odd_lor}

\begin{figure*}
\includegraphics[width=0.95\columnwidth]{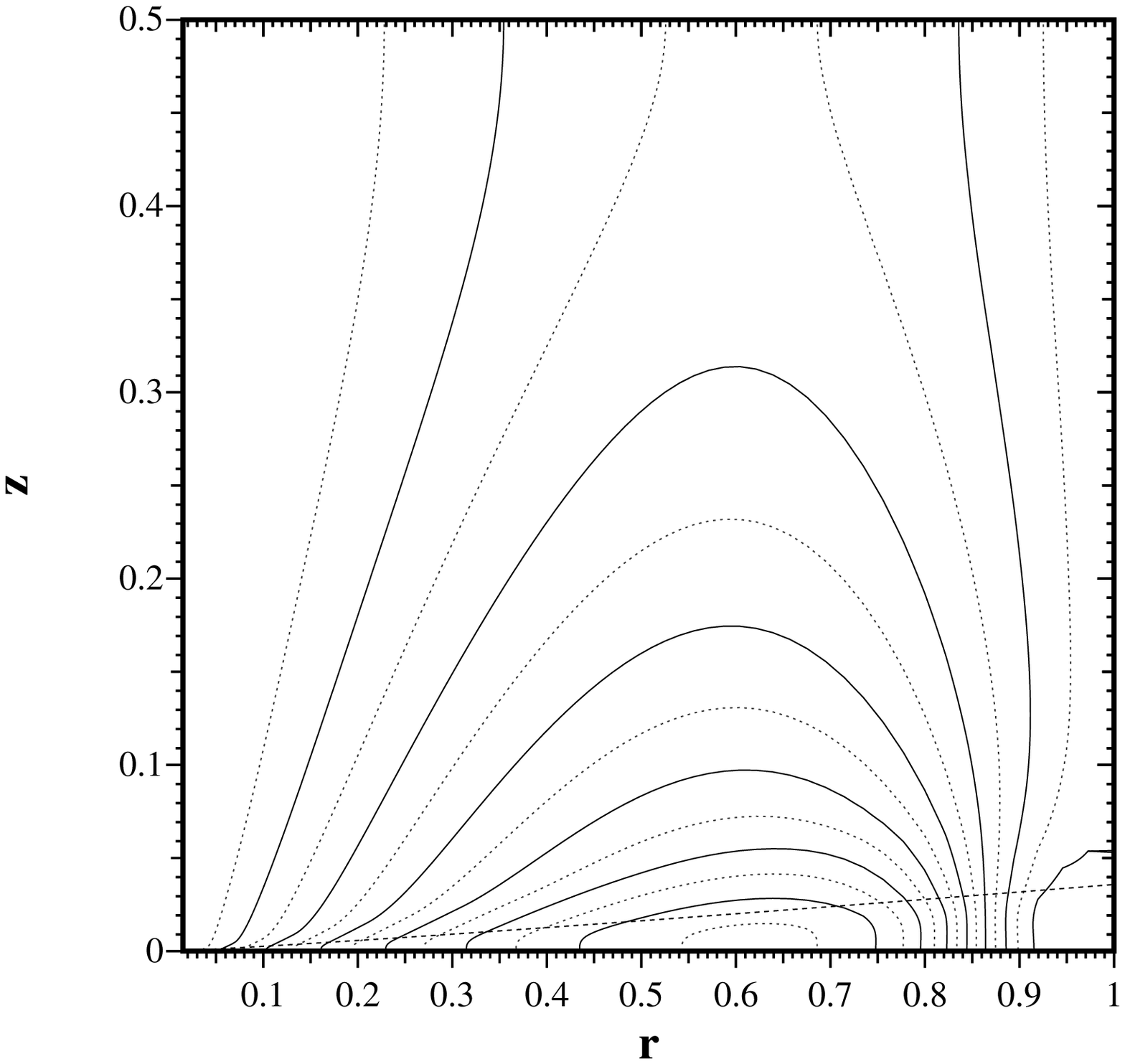} 
\includegraphics[width=0.95\columnwidth]{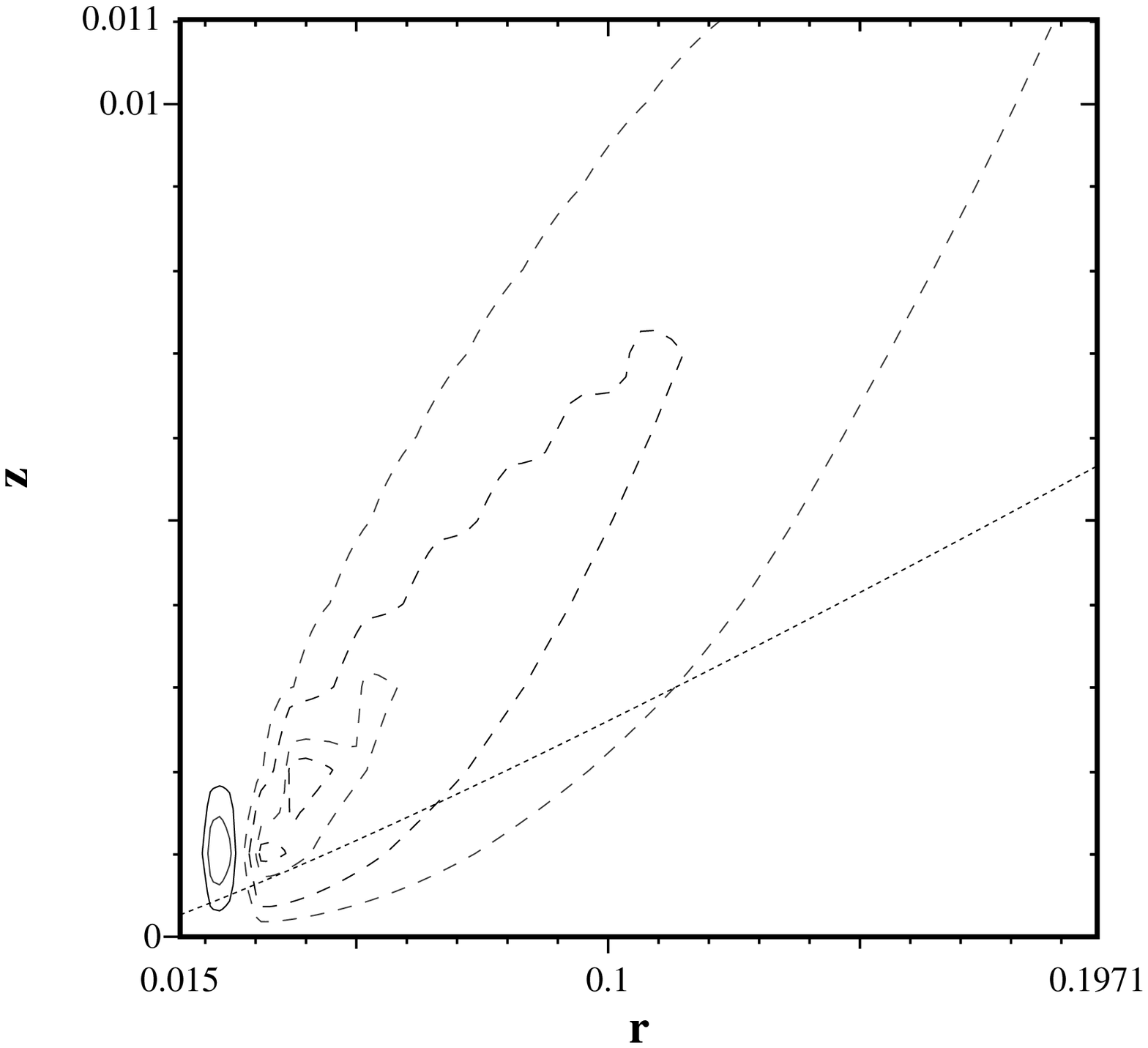} 
\caption{Global solution (left, contours of the poloidal field) and a magnified view of the solution 
near $r=r_{in}$ (right, contours of toroidal field), for the slightly 
supercritical odd parity model with $f_{\rm v} = 1.0$ and negative $R_\alpha$ -- model l55. 
The maximum value of $B_\phi$ is approximately $6.4 \times 10^4$\,G, and the contour spacing is $1.6 \times 10^4$\,G.
The dashed curve shows the disc boundary. }
\label{B_l55} 
\end{figure*}

\begin{figure}
\includegraphics[width=0.95\columnwidth]{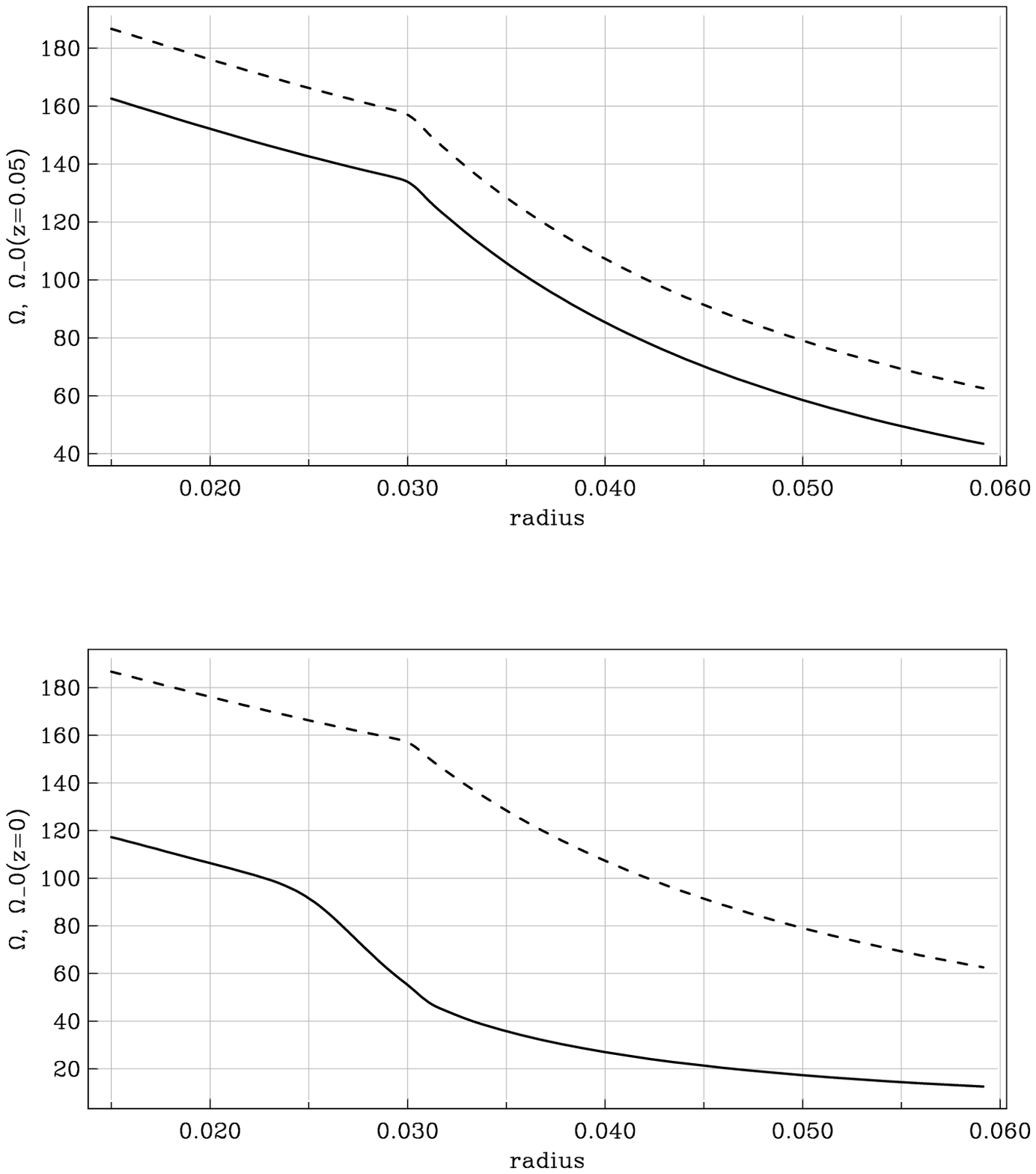}
\caption{Radial dependence of the modified angular velocity (solid)
and the initial angular velocity (broken)  at $z=0$ at $z=0.05$ 
(broken) at small radii for model l28.  }
\label{omeg_rad_l55}
\end{figure}

We also made a superficial study of models with negative $R_\alpha$
and odd parity, corresponding to domination of left-handed vortices in the northern 
hemisphere. This appears unrealistic according to contemporary 
understanding of the origin of the $\alpha$-effect, 
but it is difficult to exclude this possibility completely.
We took initially $f_{\rm v}=1$, and the poloidal and toroidal fields are shown in
Fig.~\ref{B_l55} -- model l55. The saturated field is steady. 
The dependence of angular velocity on radius in the inner
disc is shown in Fig.~\ref{omeg_rad_l55},
and
the modulation of angular velocity in the inner disc is small -- also
(Fig.~\ref{omeg_rad_l55}). Fields are not so strongly concentrated to the inner disc.  (With prescribed quadrupolar parity and $R_\alpha<0$, dynamo 
excitation occurs at significantly larger dynamo numbers: it is clear that 
with negative values of $R_\alpha$ even parity solutions are preferably excited.)
With $f_{\rm v}=1.167$, there is a transition to temporally
unsteady solutions with marked modulation of the angular velocity in the inner disc.
Accretion disc dynamo problems with negative $R_\alpha$ were addressed 
by Rekowski et al. (2000) and here we broadly confirm these results.

\section{Discussion}
\label{disc}

Simple models of magnetic field excitation in accretion discs and  the
corresponding dynamo driven magnetic 
configuration presented above show to some extent a 
similarity between accretion disc and galactic 
dynamos; however there are also important differences between them.

At least in the framework of the basic accretion disc model that uses a standard
hydrodynamical model for the structure of
discs, for smaller dynamo numbers we obtain a dynamo excitation 
with a steady magnetic field 
configuration with 
the axisymmetric magnetic field concentrated near the central plane of the disc and maximal near its inner  
boundary.
The field rapidly decays above and below the disc. The configuration weakly depends on the most prominent 
parameters responsible for accretion disc hydrodynamics, such as accretion rate, i.e. infall velocities.  

At first sight it might look rather unexpected that the 
steady magnetic field strength is approximately equal to the equipartition field strength. 
This is a natural consequence of the supercriticality of the dynamo. For example Shukurov (2004) gives the estimate 
$$B^2_{\rm steady} \sim  4\pi\rho V_t \Omega l [|D/D_cr|-1]^{1/2}$$
for the saturation field strength.
With $\eta=(l  V_t)/3$ we get
$B^2_{\rm steady}\apropto  12\pi\rho\eta\Omega$
giving a strong radial dependence
$r^{-21/8}Q(r)^{17/20}$. 
    
We can identify the parameter which could be responsible for substantial deviations from the basic model. It is the ratio 
of the r.m.s. turbulent velocity to the sound speed. Because the turbulence is barely supersonic, the ratio 
$\sigma_v$ is expected to be lower then unity. If the ratio $\sigma_v \approx 0.3$ (r.m.s. turbulent velocity is 3 
times lower then the sound speed) the magnetic configuration becomes unsteady,
and has irregular oscillations with 
time scale of about 0.2 diffusion times. 
However 
magnetic field is still concentrated at the equatorial plane. Such behaviour is unknown for galactic dynamos, maybe 
because of the lack of interest in and immediate physical relevance of such  
behaviour. An additional point is that we only have snapshots of actual galactic magnetic fields.

If we accept that the r.m.s. turbulent velocity is about one order of magnitude lower than the sound speed then, at least in the absence of feedback on the rotation curve from the Lorentz force, the 
magnetic configuration departs from strict quadrupolar symmetry and the
magnetic field maxima are found near the upper 
and lower boundaries of the disc, rather at the central disc plane. 
There are again irregular 
oscillations with dimensionless time scale slightly smaller than 
in the previous case.  Such magnetic configurations are 
unknown for galactic dynamos.  The key point here appears to be that
galactic turbulence is believed to have r.m.s. 
velocity close to the sound speed. 

Taking into account the possibility that the nature of the mirror asymmetry of accretion disc turbulence is more 
complicated then just a straightforward action of Coriolis force in a stratified medium, we here considered what 
happens if 
this mirror asymmetry has sign opposite to the conventional expectation. 
We show in Fig.~7 the field when
$f_{\rm V}=0.25$. Then the dynamo driven magnetic field is steady with    
dipolar symmetry and the magnetic field maxima (in practice, maxima of the toroidal field which has to vanish at the 
disc plane) are located above and below the disc.  

We have also made a preliminary study of the effects of the Lorentz
force arising from the large-scale magnetic field on the rotational velocities.
Except in the innermost disc, these effects are small. Generally, the overall 
conclusions are little altered. The irregular behaviour begins to occur
at slightly larger formal values of the dynamo numbers, but given the inherent uncertainty
in the modelling, including the change in definition of $\alpha(r,z)$, 
this is hardly physically significant. 

We remphasize here that the models with $f_{\rm v}<1$ correspond formally 
at least to the physically unrealistic case of supersonic turbulence in the
disc. These include most of the steady solutions. 
On the other hand, solutions with 
$f_{\rm v}>1$, corresponding formally to subsonic turbulence, are
more-or-less invariably irregular, 
both without and with the inclusion of the Lorentz force feedback on the 
rotation. However, although  we are unable to follow the latter class of 
solutions satisfactorily there are strong indications that significant 
modifications of the rotation law will be present. In turn this means that
to follow the evolution of these solutions self consistently, real time
updating of the disc model would be required. 

We note that numerical implementation even for our the simple models is relatively demanding of computer 
facilities. Making a crude estimate of what would be required for a direct 
numerical simulation of the type of magnetic 
configurations discussed, assuming as a quite modest requirement
that simulation of a box of the size $H \times 
H \times H$ would require $100^3$ mesh point and that the disc aspect ratio is $10^{-2}$, we conclude 
that simulation of magnetic field equation in the whole disc would require  about $2 \times 10^{11}$ equations, as well as
equations for hydrodynamic quantities, description of the disc environment, etc. A quite moderate 
estimate suggests that such straightforward approach would be a very 
complicated undertaking and investigation of
simple mean-field models  such as described above as an adjunct 
to direct numerical simulation looks reasonable. 
Of course, the simple 
models have specific limitations as well. In particular, it is difficult to believe that we take into account even 
the more 
important, not to say all, instabilities which result in irregular oscillation. 

In the framework of this paper we assume that the accretion disc is thin. This does not include all options 
discussed in accretion disc studies and we recognize  the possibility that magnetic field generation occurs in the 
central part of a thick disc. A natural expectation from dynamo studies is that then the magnetic field 
configuration will be an oscillating dipole with magnetic field concentrated far above and below the disc central plane; 
however verification of this expectation would require separate modelling which considers quasispherical configurations 
of dynamo drivers. This is obviously beyond the scope of this paper.

Of course, a detailed comparison of the above results with available observations of cataclysmic variables deserves 
a special investigation. At least, however, magnetic configurations obtained in the basic model appear coherent in the 
framework of the standard Shakura \& Sunyaev (1973) theory.

\subsection{Possible astrophysical applications}

We now discuss in more detail the oscillations in total magnetic energy present
in more supercritical dynamo regimes.
 Much of this discussion is based on the models without Lorentz force,
discussed in Sect.~\ref{2Dsoln}: the numerical limitations mentioned
make it difficult to compute these models for very supercritical values.
However indications are that the basic nature of the results
will not change significantly, within the intrinsic limitations and uncertainties
of the modelling.

The  corresponding  characteristic time scale of the oscillations is about 0.2-0.4 of  
the diffusion time scale. As the total (thermal plus magnetic) energy in the
disc is conserved,  the thermal energy in the disc must oscillate with the same characteristic time.
Thus  the oscillations of the magnetic energy could lead to 
quasi-periodic oscillations  (QPOs) of the accretion disc luminosity.

Our computations demonstrate that the magnetic energy is concentrated near the inner radii of accretion discs.
The maximum radiation flux also emerges at this accretion disc region. Therefore, we can expect that
QPO character times could be close to 20-40\% of the diffusion time scale characteristic of the inner parts of the disc.  

For the inner parts of the accretion discs around white dwarfs the corresponding characteristic time is
a few tens of seconds (see Eq.\,(\ref{ts})).
The coherent oscillations of high-degree with similar periods observed during dwarf nova outbursts are well known. 
They have periods 7-40 s and are known as dwarf nova oscillations (DNOs) (see Warner 2003). 
Our accretion discs have the high accretion rate typical of dwarf novae during outbursts.
Therefore, we can suggest that the magnetic energy oscillations found are connected with DNOs and could
explain them. 

Most of the existing models suggested to explain DNOs are  connected with various instabilities in the 
optically thick boundary layer between the accretion disc and the white dwarf 
(Godon 1995, Collins et al. 1998, Popham 1999, Piro \& Bildsten 2004).  
 It is interesting, that the model suggested by Warner \& Woudt (2002) describes  DNOs using 
a magnetic field in the boundary layer (a low-inertia
 magnetic accretor model).

Optically thick accretion discs also  exist around much more compact objects: neutron stars and black hole candidates (low-mass X-ray binaries, LMXBs).
Quasi-periodic oscillations  of X-ray flux are observed in LMXBs with neutron stars and they
 are known as kHz QPOs (in the range 200-1200 Hz, and usually occuring in pairs, see the review by van der Klis 2000).  
We note that the diffusion time-scale at the LMXB inner disc radii, which is close to the Kepler time scale, 
 is much smaller than in cataclysmic variables, and is comparable with the observed kHz QPO  periods 10$^{-2}$ - 10$^{-3}$ s.
 Therefore, the observed kHz QPOs can be also associated with the total magnetic energy oscillations in the accretion discs.
 
We do not propose any new model for DNOs (or kHz QPOs). 
We have just found magnetic energy oscillations in accretion 
discs due to dynamo
action, with typical time scales close to the periods of DNOs or kHz QPOs. 

There are some sources of uncertainties, which prevent us from claiming
a definitive new DNO model.
First is the nature of the $z$-dependence of the angular velocity.
The limited experimentation mentioned above suggests that this may not be too
important, but more cannot be stated. Another rather arbitrary feature of the 
model is the nature of the inner boundary conditions but, again, limited experimentation 
suggests that solutions are not very sensitive to "reasonable" choices.
     
An extension of our results to the accretion discs of black holes seems 
a natural further step for investigation. 
Treatment of the whole disc in a 2D model may be a quite challenging numerical problem, given the larger radial range and greater range of variation in disc thickness.        
Given our experiences here, further progress might be made by just modelling the innermost 
disc in a thin computational domain.

\section{Conclusions}
\label{concl}
A limitation of our models is the plausible restriction to axisymmetry. Nevertheless,
by including a realistically thin disc that extends inwards to near the 
boundary of the central accreting object we have demonstrated the importance
of including the whole disc in the model: radially truncated discs omit
the region of maximum magnetic field and the irregular fluctuations arise from this region, especially in the models with Lorentz force. 
We have been able to explore a range
of possibilities for the dynamo number, and have found quadrupolar
type solutions with irregular temporal oscillations that might be compared to observed rapid fluctuations (DNOs). 
These oscillations exist both in the models  with only the alpha-quenching
non-linearity and also for those with the addition of the Lorentz force feedback. 
The dipolar symmetry models with $R_\alpha<0$ (Sect.~\ref{ralphaneg}) have lobes of strong toroidal field adjacent 
to the rotation axis that could be relevant to jet launching phenomena.
In short, we have explored and extended the solutions known for thin accretion discs around compact  objects.

 \begin{acknowledgements} 
VS thanks DFG for financial support (grant  WE 1312/48-1).  
\end{acknowledgements} 

\section*{Appendix}

We recall here some definitions associated with decomposition of dynamo generated magnetic configurations into symmetry types with respect to the disc plane.
Let $z=0$ be the central plane of an accretion disc in a Cartesian 
reference frame $x, y, z$.
Define the even and odd parts of a scalar $S(x,y,z)$ by
$S_{\rm even}=(S(x,y,z)+S(x,y,-z))/2$  and $S_{\rm odd}=(S(x,y,z)-S(x,y,-z))/2$.
If ${\bf B} (x, y, z)$ is a magnetic field, then define
$E_{\rm even}=\int(B_{x,{\rm even}}^2+B_{y, {\rm even}}^2+B_{z,{\rm odd}}^2)/(8\pi)dV$ and correspondingly
 $E_{\rm odd}=\int(B_{x,{\rm odd}}^2+B_{y, {\rm odd}}^2+B_{z,{\rm even}}^2)/(8\pi)dV$. The mixing of odd and even parts in these expressions is a 
consequence of an alpha-effect dynamo with $\alpha$-coefficient that is odd with
respect to the plane $z=0$. 
The integrals are in principle taken over the whole space and in practice over the computational domain.

\end{document}